\documentclass{aa}
\usepackage{graphicx}
\usepackage{txfonts}
\usepackage{supertabular}
\usepackage{longtable}
\usepackage{xcolor}
\usepackage{caption}
\usepackage{siunitx}
\definecolor{dark-gray}{gray}{0.35}
\begin{document}

   \title{Singly- and doubly-deuterated formaldehyde in massive star-forming regions}

   \author{S. Zahorecz
          \inst{1,2}
          \and
          I. Jimenez-Serra
          \inst{3}
          \and
          L. Testi
          \inst{4}
          \and
          K. Immer
          \inst{5}
		  \and
          F. Fontani
          \inst{6}
		  \and
          P. Caselli
          \inst{7}
          \and
          K. Wang
          \inst{8}
          \and
          T. Onishi
          \inst{1}
          }

   \institute{Department of Physical Science, Graduate School of Science, Osaka Prefecture University, 1-1 Gakuen-cho, Naka-ku, Sakai, Osaka 599-8531, Japan\\
    \email{s.zahorecz@p.s.osakafu-u.ac.jp}
            \and
 Chile Observatory, National Astronomical Observatory of Japan, National Institutes of Natural Science, 2-21-1 Osawa, Mitaka, Tokyo 181-8588, Japan
            \and
Centro de Astrobiolog\'ia (CSIC/INTA), Ctra de Torrej\'on a Ajalvir, km 4, E-28850 Torrej\'on de Ardoz, Spain
            \and
European Southern Observatory, Karl-Schwarzschild-Str. 2, 85748, Garching bei M\"unchen, Germany
        \and
		Joint Institute for VLBI ERIC (JIVE), Postbus 2, NL-7990 AA Dwingeloo, the Netherlands
		\and
        INAF-Osservatorio Astrofisico di Arcetri, L.go E. Fermi 5, 50125 Firenze, Italy
 		\and
        Max-Planck-Institut f\"ur Extraterrestrische Physik, Giessenbachstrasse 1, 85748 Garching bei M\"unchen, Germany
		\and
		Kavli Institute for Astronomy and Astrophysics, Peking University, 5 Yiheyuan Road, Haidian District, Beijing 100871, China
    }

   \date{Received ; accepted }

 
  \abstract
  {Deuterated molecules are good tracers of the evolutionary stage of star-forming cores. During the star formation process, deuterated molecules are expected to be enhanced in cold, dense pre-stellar cores and to deplete after protostellar birth. }
   {In this paper we study the deuteration fraction of formaldehyde in high-mass star-forming cores at different evolutionary stages to investigate whether the deuteration fraction of formaldehyde can be used as an evolutionary tracer.}
   {Using the APEX SEPIA Band 5 receiver, we extended our pilot study of the $J$=3$\rightarrow$2 rotational lines of HDCO and D$_2$CO to eleven high-mass star-forming regions that host objects at different evolutionary stages. High-resolution follow-up observations of eight objects in ALMA Band\,6 were performed to reveal the size of the H$_2$CO emission and to give an estimate of the deuteration fractions HDCO/H$_2$CO and D$_2$CO/HDCO at scales of $\sim$6'' (0.04-0.15 pc at the distance of our targets).}
   {Our observations show that singly- and doubly deuterated H$_2$CO are detected toward high-mass protostellar objects (HMPOs) and ultracompact \ion{H}{II} regions (UC~\ion{H}{II} regions), the deuteration fraction of H$_2$CO is also found to decrease by an order of magnitude from the earlier HMPO phases to the latest evolutionary stage (UC~\ion{H}{II}), from $\sim$0.13 to $\sim$0.01. We have not detected HDCO and D$_2$CO emission from the youngest sources (high-mass starless cores, HMSCs).}
   {Our extended study supports the results of the previous pilot study: the deuteration fraction of formaldehyde decreases with evolutionary stage, but higher sensitivity observations are needed to provide more stringent constraints on the D/H ratio during the HMSC phase. The calculated upper limits for the HMSC sources are high, so the trend between HMSC and HMPO phases cannot be constrained.}


   \maketitle 

%

\section{Introduction}
Deuterium fractionation, and how it changes as star-formation proceeds, is a powerful tool to establish the evolutionary stage and age of star-forming cores. The formation of deuterated molecules is favoured at low temperatures (T $\leq$ 20~K) and at high densities (n $\geq 10^4$~cm$^{-3}$) in molecular cloud cores, at the ideal conditions for CO freeze out \citep{Caselli1999, Caselli2002a}. 
However, deuterated molecules are also observed with relatively high abundances in warmer environments, e.g. in circumstellar disks and in hot cores \citep[e.g. ][]{Salinas2017, Sakai2018}. Different chemical processes contribute to the deuteration enrichment at different temperatures. At low (T=10-20\,K) temperatures the dominant gas-phase process is H$_3^+$+HD$\rightleftharpoons$H$_2$D$^+$+H$_2$. At high temperatures (T=30-50\,K), however, the CH$_3^+$+HD$\rightleftharpoons$CH$_2$D$^+$+H$_2$ and C$_2$H$_2^+$+HD$\rightleftharpoons$C$_2$HD$^+$+H$_2$ reactions become important \citep[see detailed gas-phase deuteration models by e.g.][]{Roueff2007, Roueff2013}.
Reactions on the grain surfaces can also influence the D/H ratio via low-temperature surface reactions during the pre-stellar/starless phase that enrich the D/H ratio of molecules in the solid phase. These molecules can then be desorbed into the gas phase during the warm protostellar phase \citep[e.g.][]{Brown1989a, Brown1989b, Tielens1983}. 

We can characterize the deuteration fraction, $D_\mathrm{frac}$, as the relative abundance between a species containing a deuterium atom as compared to the same species containing a hydrogen atom. It is thus expected that $D_\mathrm{frac}$ is enhanced in cold and dense pre-stellar cores, and it should then decrease after protostellar birth, when the young stellar object heats up the central region of the core \citep{Caselli2002b}. Several studies \citep[see eg. ][]{Crapsi2005, Caselli2008, Emprechtinger2009} have confirmed this theoretical scenario for low-mass star forming cores, with the observations of deuterated species produced by gas-phase reactions (e.g. H$_2$D$^+$ and N$_2$D$^+$). Recent studies have shown that high $D_\mathrm{frac}$ values are also typical for high-mass star-forming cores and that the $D_\mathrm{frac}$ of some species could be an evolutionary tracer also in the intermediate and high-mass regime \citep[e.g. ][]{Busquet2010, Fontani2011, Fontani2015, Sakai2012, Zahorecz2017}. 

Recent theoretical studies \citep[e.g.][]{Bovino2017, Kortgen2018, Hsu2021} have shown that the evolution of the deuterium enrichment depends on several factors, such as the initial physical conditions, grain size distribution, depletion fraction and cosmic ray ionization rate. Observational studies \citep[e.g.][]{Crapsi2005, Emprechtinger2009, Fontani2011, Fontani2015, Trevino2014} have confirmed that the deuterium chemistry is a good chemical clock for both low-mass and high-mass star-forming cores. However, in order to have an accurate estimate of the evolutionary stage of the source, it is necessary to use as many different molecular tracers as possible since the derived D$_\mathrm{frac}$ may differ due to variations in the chemical evolution of the different species \citep[see e.g.,][]{Trevino2014}.

We can distinguish the following evolutionary stages in high-mass star-forming regions: HMSCs - high-mass starless cores, HMPOs - high-mass protostellar objects and UC~\ion{H}{II} regions - ultracompact \ion{H}{II} regions \citep[e.g. ][]{Beuther2007,Tan2014}. \citet{Gerner2014} derived chemical age of $\sim$6$\times$10$^4$, $\sim$4$\times$10$^4$ and $\sim$10$^4$ yrs for the HMSC, HMPO and UC~\ion{H}{II} stage, respectively. Their derived total timescale for the high-mass star-formation is consistent with theoretical estimates of $\sim$10$^5$ yrs. Based on the study of several deuterated species in 27 massive cores in different evolutionary stages from HMSCs to UC~\ion{H}{II} regions, \citet{Fontani2011, Fontani2015} found that species formed exclusively in the gas (N$_2$H$^+$) showed different evolutionary trends from those formed partially (NH$_3$) or totally (CH$_3$OH) on grain mantles. 
The $D_\mathrm{frac}$(N$_2$H$^+$) is higher in the HMSC phase and it drops by about an order of magnitude during the HMPO and UC~\ion{H}{II} stages, while deuterated methanol is detected only towards HMPOs and externally heated HMSCs. Therefore, high $D_\mathrm{frac}$(N$_2$H$^+$) is a good indicator of the initial conditions in starless/pre-stellar cores, while high $D_\mathrm{frac}$(CH$_3$OH) values are a good probe of the earliest protostellar phases \citep{Fontani2015}.
$D_\mathrm{frac}$(NH$_3$) does not show significant difference across the different evolutionary stages, and thus, there is no dominant formation pathways (gas-phase vs. grains) for NH$_3$ and its deuterated forms.

Like NH$_3$, formaldehyde (H$_2$CO) can also be produced both in the gas phase and on grain surfaces. The two main pathways for the production of H$_2$CO involve CH$_3^+$ in the gas phase \citep[see][]{Yamamoto2017} and multiple hydrogenation of CO on the ices \citep[see e.g.][]{Roberts2007}. Its gas phase pathway is similar to the one of N$_2$H$^+$, but it can also occur at warmer temperatures \citep[30-50\,K, see][]{Parise2009}. The ice phase formation route of H$_2$CO is similar to that of CH$_3$OH and their deuterated forms. Laboratory work and observational studies of low-mass protostars suggest an important contribution from grain surface chemistry to the production of H$_2$CO \citep[][]{Watanabe2005, Roberts2007, Bergman2011}. However the relative importance of the grain surface versus gas-phase formation routes remains unclear. Furthermore, only a handful D$_2$CO detections in high-mass star-forming regions are available \citep[D$_2$CO has been firmly detected toward the Orion Compact Ridge, NGC\,7129-FIRS\,2, AFGL5142, IRAS05358+3543 and tentatively toward the MonR2 ultra-compact \ion{H}{II} region;][]{Fuente2005, Trevino2014, Turner1990, Zahorecz2017}. 

All these initial results suggest that H$_2$CO and its deuterated species form mostly on grain surfaces, but more observations carried out with interferometers (i.e. at high-angular resolution) are needed to separate the HDCO and D$_2$CO emission originating from the small and dense cores and to get a more detailed view of the formaldehyde deuteration in high-mass star-forming regions.

In this paper, we report our findings of a search for HDCO and D$_2$CO emission toward an extended sample of high-mass star-forming regions at different evolutionary stages, using APEX and ALMA telescopes. The source sample and the observations are described in Section 2, while our results are presented in Section 3. In Section 4, we discuss our results and put them in context with respect to previous findings in low-mass and high-mass star-forming regions. In Section 5, we summarize our conclusions.

\section{Observations and target selection}
In this section, we describe our selected sources and the single-dish and interferometric observations we have carried out for our study. 
\subsection{Selected targets}
The selected sources are extracted from the sample of \citet{Fontani2011}, for which the deuteration fraction of other molecules (e.g. N$_2$H$^+$, CH$_3$OH and NH$_3$) has been measured \citep{Fontani2015}. We have selected the brightest sources from each evolutionary class, which are observable with the Atacama Pathfinder EXperiment (APEX\footnote{This publication is based on data acquired with the Atacama Pathfinder Experiment (APEX). APEX is a collaboration between the Max-Planck-Institut f\"ur Radioastronomie, the European Southern Observatory, and the Onsala Space Observatory.}) telescope from the southern hemisphere. Table \ref{table:basic_data} reports the observed central coordinates, velocities, distances, kinetic temperatures and source evolutionary classification for our targets. The sample is divided into three high-mass starless cores (HMSCs), three high-mass protostellar objects (HMPOs), two sources containing HMSC and HMPO (which would represent an intermediate stage between HMSC and HMPO) and three ultracompact \ion{H}{II} regions, see details below. The source coordinates have been centred at interferometric mm/cm continuum peaks or high-density gas tracer peaks (e.g. NH$_3$ with VLA) that are separated by at least $\sim$10$\arcsec$ from other cores. The APEX beam is $\sim$34'' at 190\,GHz. Therefore, for AFGL5142 and IRAS05358+3543, multiple cores at different evolutionary stages are covered within the same pointing. We categorize these two sources as a transitional stage between HMSCs and HMPOs in our work.

\begin{table*}[!ht]
\centering
\begin{tabular}{c c c S[table-format=2.1] S[table-format=1.2] c c}
\hline
Name & RA &  DEC & v$_\mathrm{LSR}$ & {d} & T$_\mathrm{kin}$ & $^{12}$C/$^{13}$C\\ 
& [hh:mm:ss.s] &  [dd:mm:ss] & [km/s] & [kpc] & [K] &\\ 
\hline
\multicolumn{6}{c}{HMSC} \\ \hline
G028-C1 & 18:42:46.9 & -04:04:08 & +78.3 & 5.0 & 17 & 37.3\\
G034-F2 & 18:53:16.5 & +01:26:10 & +57.7 & 3.7 & - & 45.4\\
G034-G2 & 18:56:50.0 & +01:23:08 & +43.6 & 2.9 & - & 50.4\\
\hline
\multicolumn{6}{c}{HMSC/HMPO} \\ 
\hline
AFGL5142 & 05:30:48.0 & +33:47:54 & -3.9 & 1.8 & 25$^*$ & 57.2\\ 
IRAS05358+3543& 05:39:13.0 & +35:45:51 & -17.6 & 1.8 & 35$^*$ & 57.2\\
\hline
\multicolumn{6}{c}{HMPO} \\ 
\hline
18089-1732 & 18:11:51.4 & -17:31:28 & +32.7 & 3.6 & 38 & 46.0\\
18517+0437 & 18:54:14.2 & +04:41:41 & +43.7 & 2.9 & - & 50.4\\
G75-core & 20:21:44.0 & +37:26:38 & +0.2 & 3.8 & 96 & 44.8\\
\hline
\multicolumn{6}{c}{UC~\ion{H}{II}} \\ \hline
G5.89-0.39 & 18:00:30.5 & -24:04:01 & +9.0 & 1.3 & - & 60.4\\ 
I19035-VLA1 & 19:06:01.5 & +06:46:35 & +32.4 & 2.2 & 39 & 54.7\\
19410+2336 & 19:43:11.4 & +23:44:06 & +22.4 & 2.1 & 21 & 55.3\\
\hline
\end{tabular}
\normalsize
\caption{List of the observed sources and their coordinates, LSR velocities, distances, kinetic temperatures, and $^{12}$C/$^{13}$C ratio at the corresponding distance. The distances and kinetic temperatures were adopted from \citet{Fontani2011}. The $^*$ symbols indicate that we use the average T$_\mathrm{kin}$ values of the HMSCs and HMPOs within our APEX beam. \label{table:basic_data}}
\end{table*}

\subsection{IRAM-30m H$_2$CO observations}
Previous observations of these objects were performed using the IRAM-30m telescope \citep[details can be found in][]{Fontani2011, Fontani2015}. The frequency setups included the 3$_{0,3}-2_{0,2}$, 3$_{2,2}-2_{2,1}$ and 3$_{2,1}-2_{2,0}$ transitions of H$_2$CO. The spectroscopic information of these transitions is shown in Table \ref{table:observed_transitions} \citep{Muller2005}. 
The spectra were obtained and calibrated in antenna temperature units, T$_A^*$ and then converted to main beam brightness temperature, T$_\mathrm{MB}$, with the following equation: T$_A^*$ = T$_\mathrm{MB} \eta_\mathrm{MB}$, where a main beam efficiency, $\eta_\mathrm{MB}$ of 0.66 was used.
The H$_2$CO line emission was detected in all sources and the measured intensities show a wide range between 0.24\,K$<T_\mathrm{MB}<$13\,K (see Fig.~\ref{figure:all_spectra}). 

\subsection{APEX SEPIA observations}
The sources were observed with the APEX SEPIA receiver \citep[Swedish-ESO PI receiver for APEX;][]{Billade2012, Belitsky2018}, as part of the ESO E-095.F-9808A, E-096.C-0484A and E097.C-0897A projects between July 2015 and June 2016. The first results based on the science verification dataset for AFGL5142, IRAS05358+3543 and G5.89-0.39 were presented in \citet{Zahorecz2017}. We carried out single-pointing observations of HDCO and D$_2$CO emission using the position switching observing mode. The J2000 central coordinates used in our observations are shown in Table\,\ref{table:basic_data}.
The observed transitions of HDCO (at 185\,GHz and 193\,GHz) and D$_2$CO (at 174.4\,GHz) are shown in Table \ref{table:observed_transitions}. The pointing was checked every 60-90\,mins, and the typical system temperatures were 150\,K. The beam size at 183\,GHz was $\sim$34$\arcsec$. The spectra were obtained and calibrated in antenna temperature units and then converted to main beam brightness temperature using $\eta_\mathrm{MB}$ of 0.68. The XFFTS spectrometer provided a velocity resolution of 0.112\,km\,s$^{-1}$, 0.059\,km\,s$^{-1}$ and 0.066\,km\,s$^{-1}$ for the H$_2^{13}$CO (at 206\,GHz), HDCO (at 185-193\,GHz) and D$_2$CO (at 174\,GHz) transitions, respectively. However, for the data analysis, we smoothed the spectra to a uniform velocity resolution of 0.5\,km/s. At this velocity resolution, we reached an rms of $\sim$ 0.01\,K with typical integration times between 32 and 48 minutes. The PWV was 0.3-1.5\,mm. The data were reduced and analysed with the Gildas software\footnote{See http://www.iram.fr/IRAMFR/GILDAS.} \citep{Pety2005}. 

\subsection{ALMA Cycle 5 observations}
We performed high-resolution single pointing observations with ALMA in Cycle 5 as part of the 2017.1.01157.S project (PI: S. Zahorecz). We used the Atacama Compact Array \citep[ACA, Morita Array;][]{Iguchi2009}. Band\,6 observations were executed for G028-C1, G34-F2, G34-G2, 18089-1732, 18517+0437, G5.89-0.39, I1935-VLA1 and 1941+2336. We used two setups to observe selected molecular lines, including H$_2^{13}$CO, and to observe the continuum emission. Table\,\ref{table:observed_transitions} contains the basic information of the observed H$_2^{13}$CO transition. In this paper, we use the continuum and H$_2^{13}$CO emission information at 213\,GHz to derive the source sizes. A detailed analysis of all the detected lines will be presented in a following paper. A continuum sensitivity of at least $\sim$20\,mJy/beam was achieved for the targets. For the three HMSC cores, a higher sensitivity of $\sim$3\,mJy/beam was achieved. For the molecular line observations, a sensitivity of 0.1\,Jy/beam was achieved at a velocity resolution of 0.086\,km/s.

\begin{table*}[tbh]
\begin{center}
\begin{tabular}{ccrccrr}
\hline \noalign {\smallskip}
Transition & Frequency & E$_u$ & Telescope& B$_{eff}$ & $\Delta$v & FWHM \\
& [MHz] & [K] & & & [km\,s$^{-1}$] & [''] \\
\hline \noalign {\smallskip}
\multicolumn{5}{l}{H$_2$CO} \\
3$_{0,3}-2_{0,2}$& 218222.192& 10.5 & IRAM & 0.62 & 0.26 & 11 \\
3$_{2,1}-2_{2,0}$& 218760.066& 57.6 & IRAM & 0.62 & 0.26 &11 \\
3$_{2,2}-2_{2,1}$& 218475.632& 57.6 & IRAM & 0.62 & 0.26 &11 \\
\multicolumn{5}{l}{H$_2^{13}$CO} \\
3$_{0,3}-2_{0,2}$& 212811.184& 10.2 & ALMA & 0.75 & 0.09 & 5 - 8\\
3$_{1,2}-2_{1,1}$& 219908.525& 22.4 & IRAM & 0.62 & 0.26 &11 \\
3$_{1,3}-2_{1,2}$& 206131.626& 21.7 & APEX & 0.72 & 0.06 & 30 \\
\multicolumn{5}{l}{HDCO} \\
3$_{0,3}-2_{0,2}$& 192893.256& 9.3 & APEX & 0.72 & 0.06 & 32 \\
3$_{1,3}-2_{1,2}$& 185307.113& 16.9 & APEX & 0.72 & 0.06 & 34 \\
3$_{2,1}-2_{2,0}$& 193907.460& 41.1 & APEX & 0.72 & 0.06 & 32 \\
3$_{2,2}-2_{2,1}$& 193391.605& 41.1 & APEX & 0.72 & 0.06 & 32 \\
\multicolumn{5}{l}{D$_2$CO} \\
3$_{0,3}-2_{0,2}$& 174413.115& 8.4 & APEX & 0.72 & 0.07 &36 \\
\hline \noalign {\smallskip}
\end{tabular}
\caption[]{Targeted transitions and beam sizes, beam coupling efficiencies and spectral resolution} of our IRAM 30m, APEX SEPIA and ALMA Band\,6 observations. {\label{table:observed_transitions}}
\end{center}
\end{table*}

\section{Results}
In this section, we present the line and continuum emission detected toward our targets. We use the ALMA Band 6 data to estimate the source sizes. Our calculations of the excitation temperature and column density of formaldehyde and its deuterated species are based on the single-dish APEX and IRAM spectra. We also present the column density ratios of the detected molecules.

\subsection{Line spectra of deuterated H$_2$CO species}
In Fig.~\ref{figure:all_spectra}, we present the H$_2$CO, HDCO and D$_2$CO 3$_{0,3}-2_{0,2}$ lines and the H$_2^{13}$CO 3$_{1,2}-2_{1,1}$ line toward our sources observed with the IRAM-30m and APEX telescopes. Six sources show multi-velocity components; we performed the calculations separately for them. Neither HDCO nor D$_2$CO emission were detected toward the HMSC sources. The HDCO lines (upper energy levels of 18.5-50.4\,K) are clearly detected in seven sources (all sources except the HMSC sources and I19035-VLA1). The D$_2$CO 3$_{0,3}-2_{0,2}$ transition is only detected clearly toward four sources: the two HMSC/HMPO sources (AFGL5142 and IRAS05358+3543), one HMPO object (18517+0437) and one UC~\ion{H}{II} region (19410+2336). We tentatively detected the same line toward one more HMPO object, G75-core and one more UC~\ion{H}{II} regions, I19035-VLA1, but further observations are needed to confirm these detections. We use only the clear detections in our calculations, and consider the rest as upper limits. 

\begin{figure*}[!ht]
  \centering
  \includegraphics[width=0.53\columnwidth]{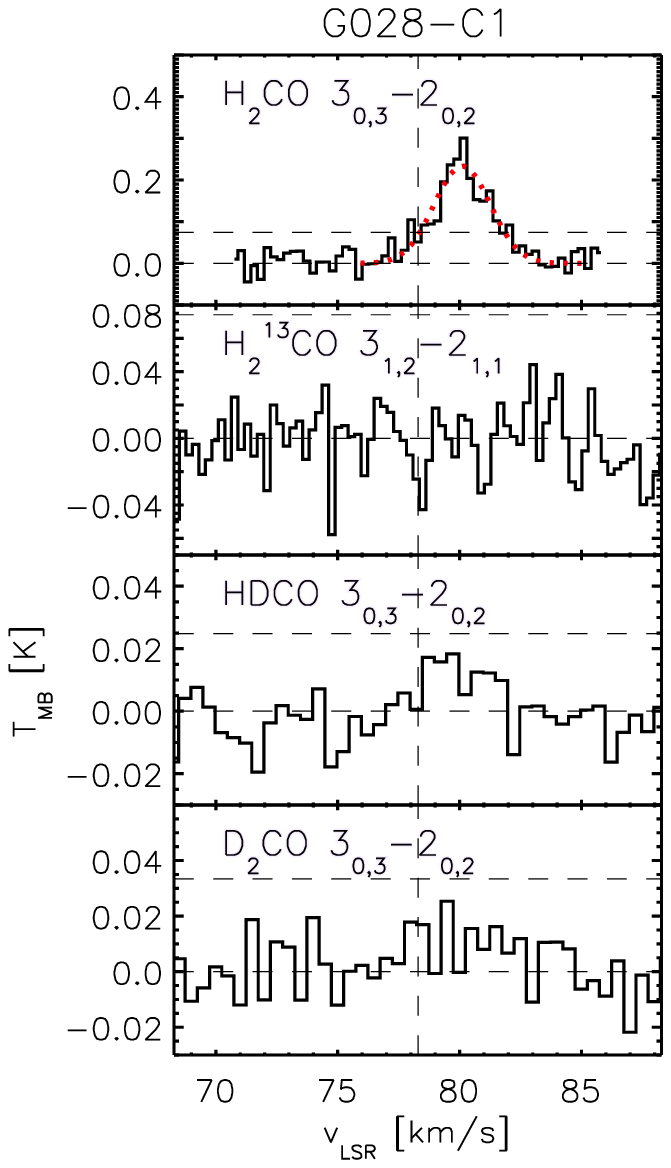}
  \includegraphics[width=0.53\columnwidth]{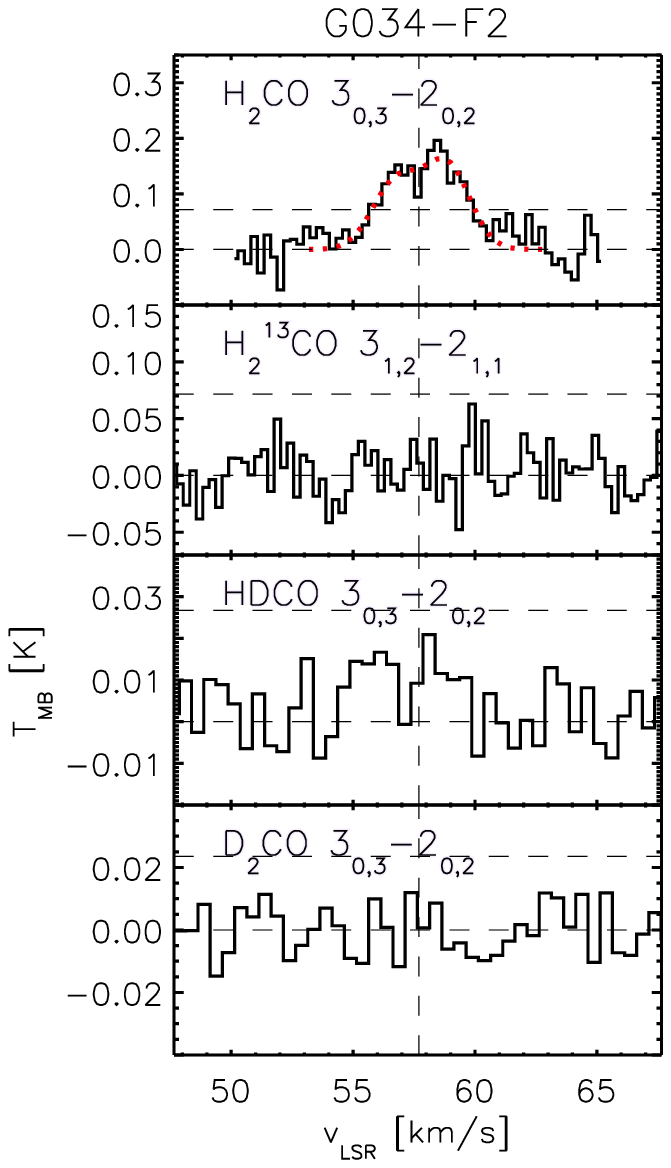}
  \includegraphics[width=0.53\columnwidth]{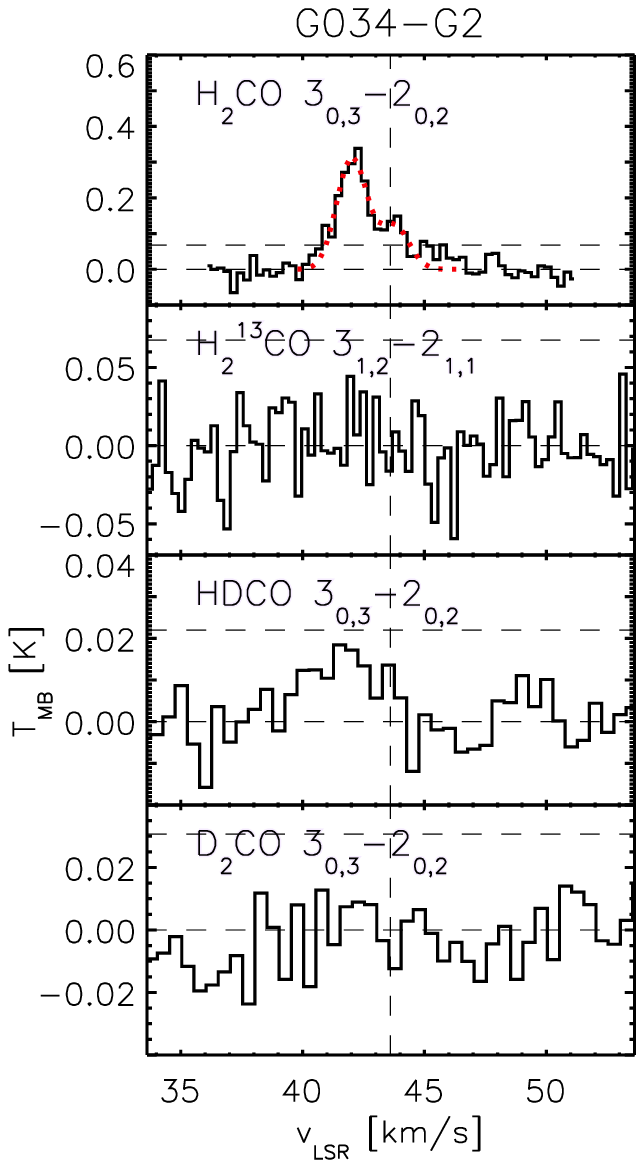}
  \includegraphics[width=0.53\columnwidth]{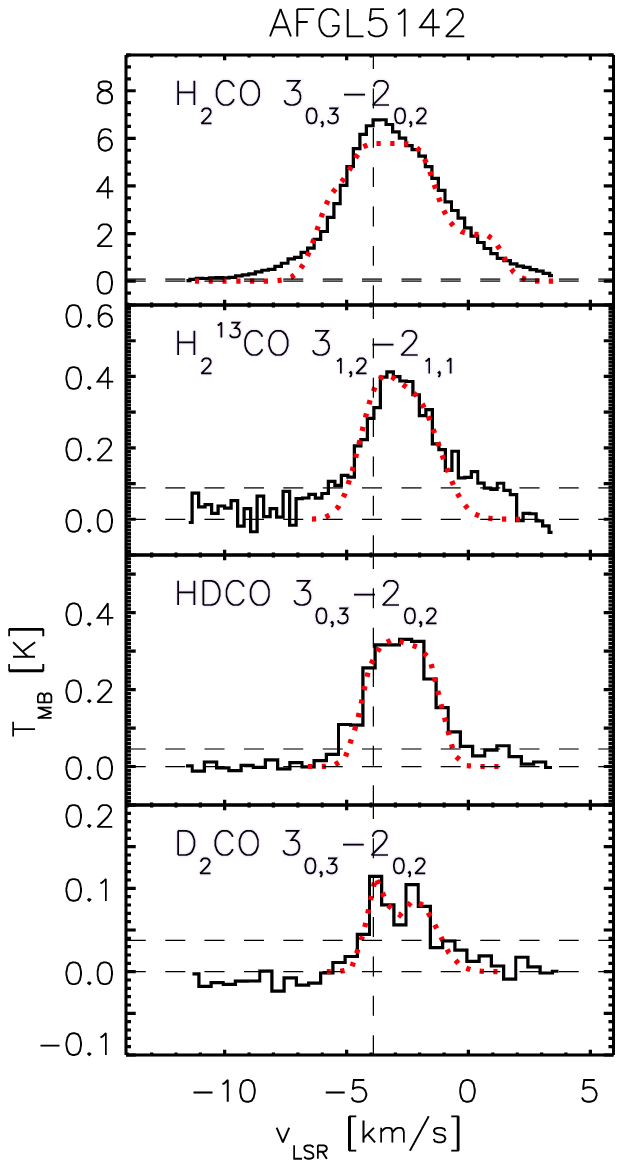}
  \includegraphics[width=0.53\columnwidth]{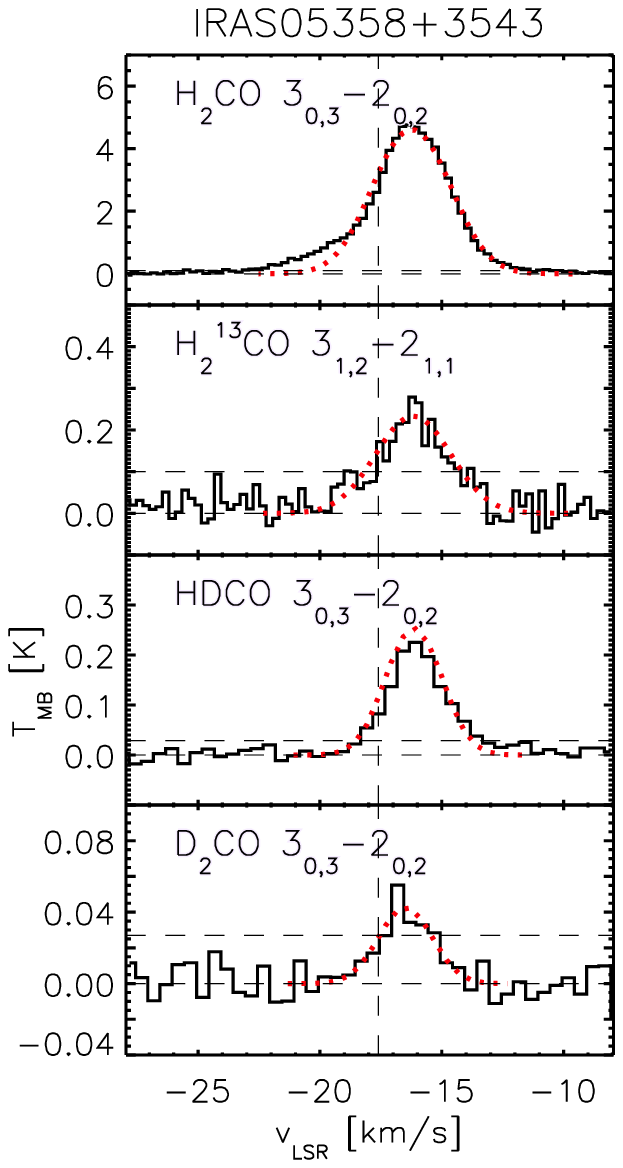}
  \includegraphics[width=0.53\columnwidth]{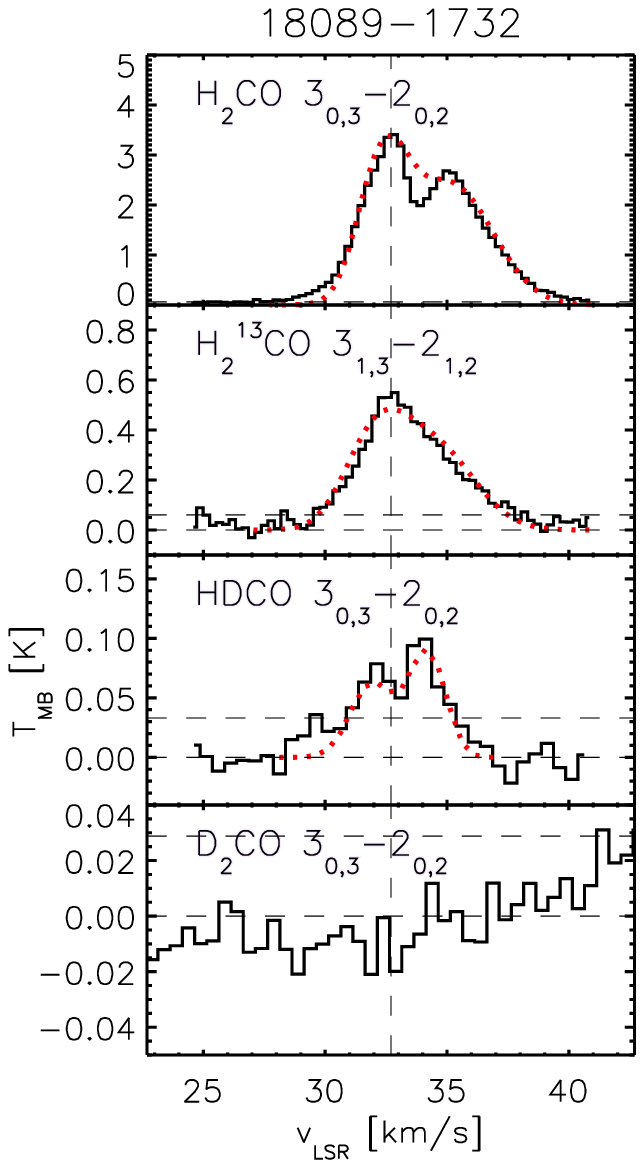}
  \caption{Spectra of the H$_2$CO 3$_{0,3}-2_{0,2}$ and H$_2^{13}$CO 3$_{1,2}-2_{1,1}$ lines (top two panels) observed with the IRAM 30m telescope \citep[][]{Fontani2011, Fontani2015} and HDCO 3$_{0,3}-2_{0,2}$ and D$_2$CO 3$_{0,3}-2_{0,2}$ lines (bottom two panels) observed with APEX SEPIA Band 5 receiver \citep[][and this work]{Zahorecz2017}. Red lines indicate the fitted spectra. Horizontal lines indicate the T$_\mathrm{MB}$=0\,K and the 3\,$\sigma$ level. Vertical line indicates the V$_\mathrm{LSR}$ for each source. The sources are sorted by evolutionary state, the order is the same as their order in Table \ref{table:basic_data}.
  \label{figure:all_spectra}}
\end{figure*}

\begin{figure*}[!ht]
\ContinuedFloat
  \centering
  \includegraphics[width=0.53\columnwidth]{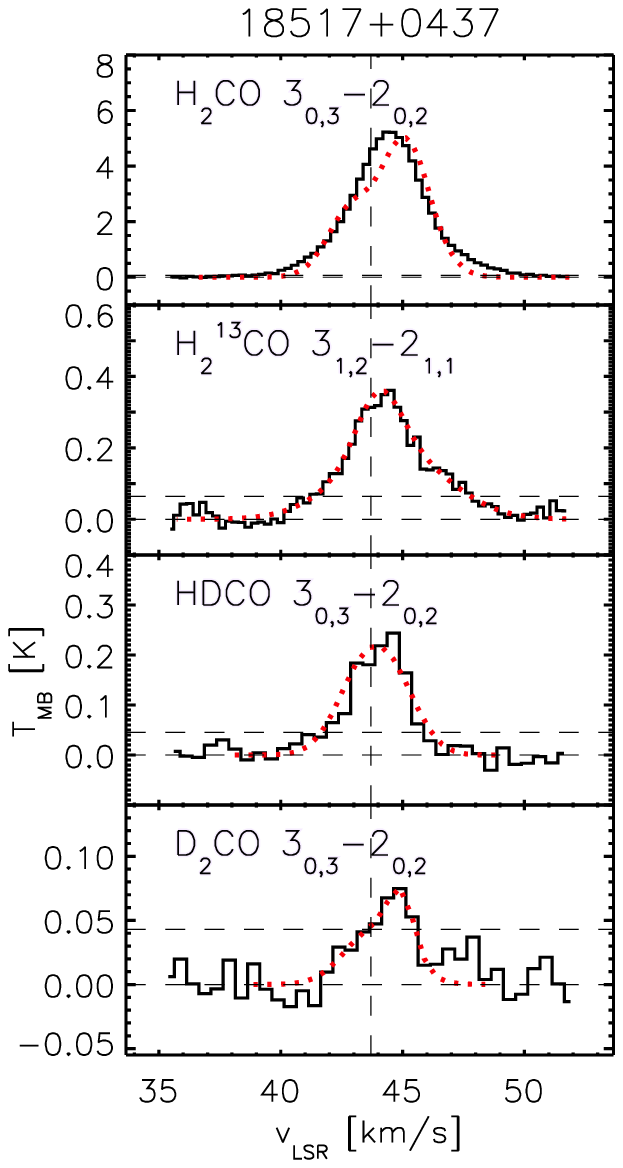}
  \includegraphics[width=0.53\columnwidth]{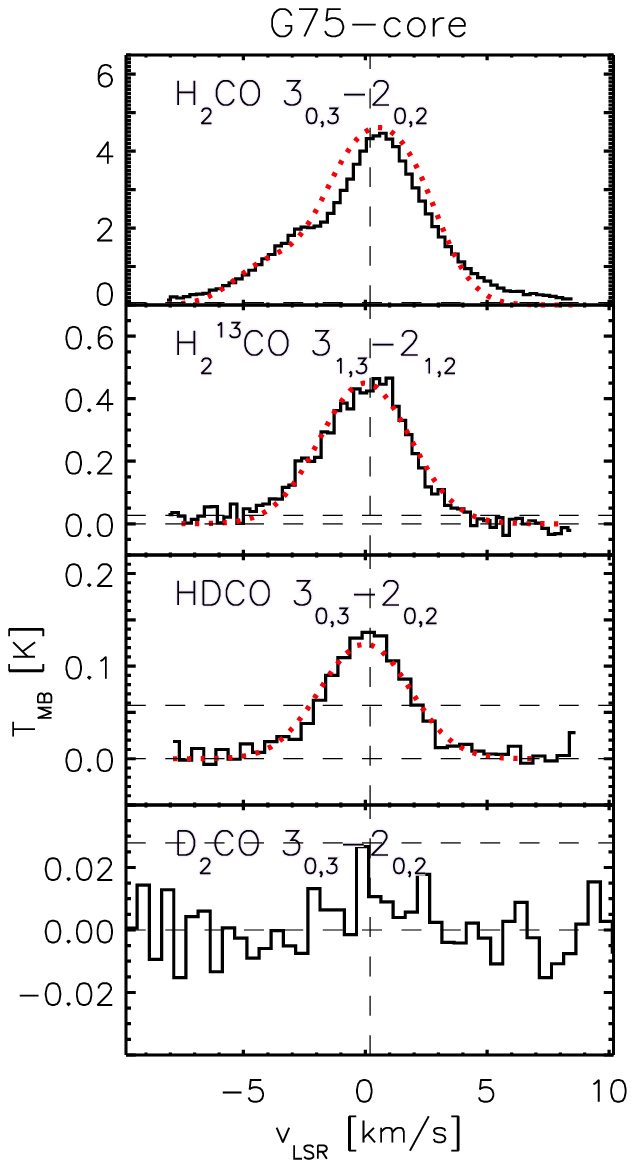}
  \includegraphics[width=0.53\columnwidth]{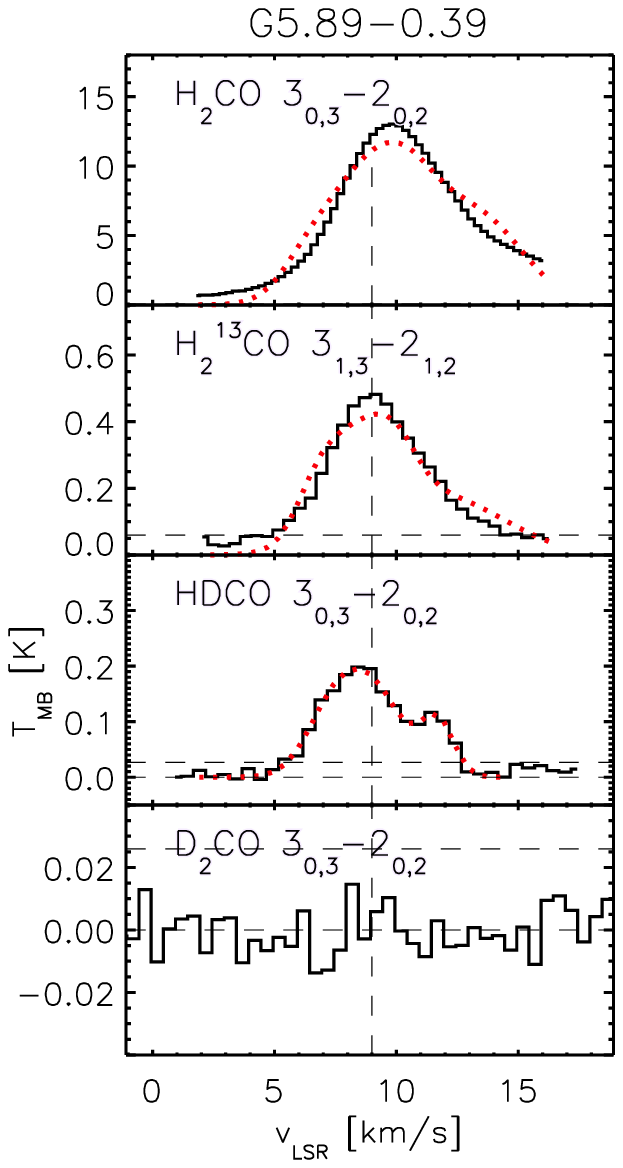}
  \includegraphics[width=0.53\columnwidth]{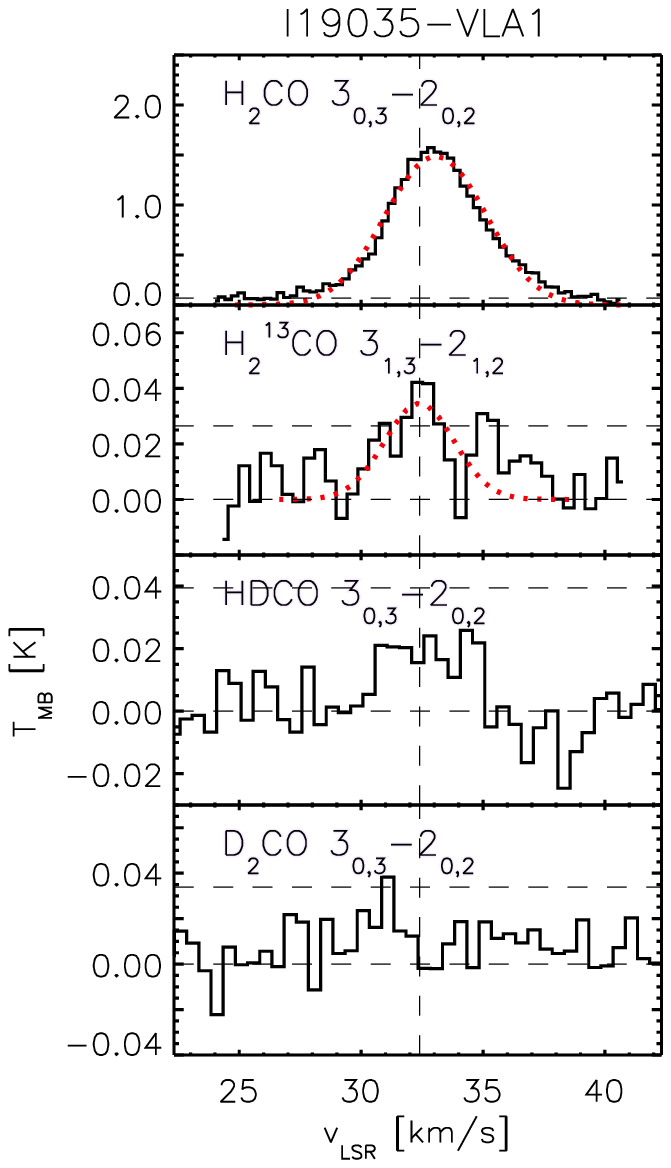}
  \includegraphics[width=0.53\columnwidth]{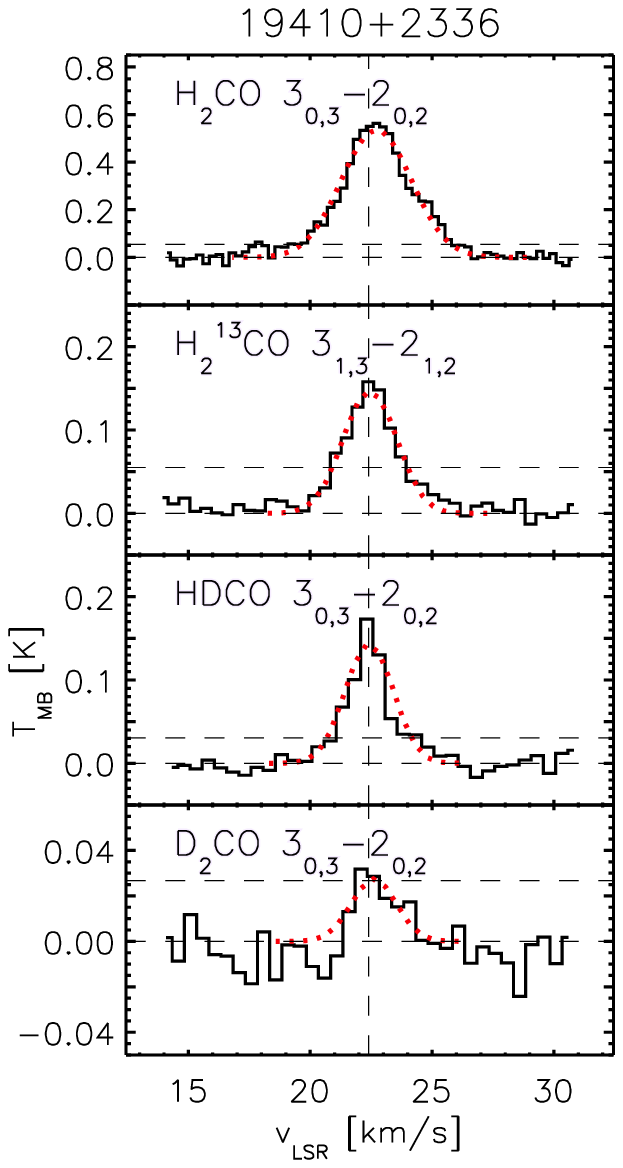}
\caption{cont}
\end{figure*}

\begin{figure*}[!ht]
  \centering
   \includegraphics[width=0.7\columnwidth]{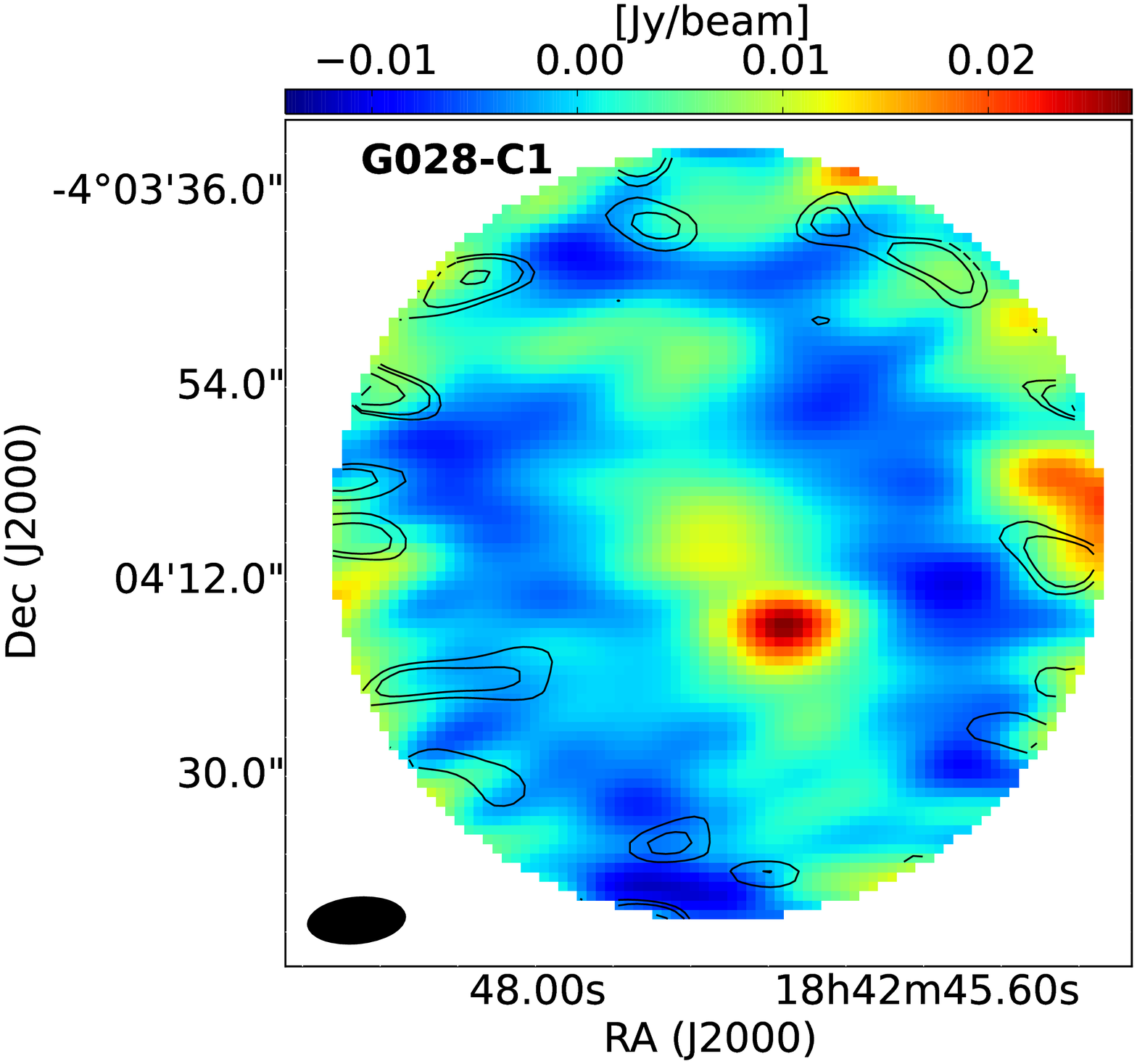}
  \includegraphics[width=0.7\columnwidth]{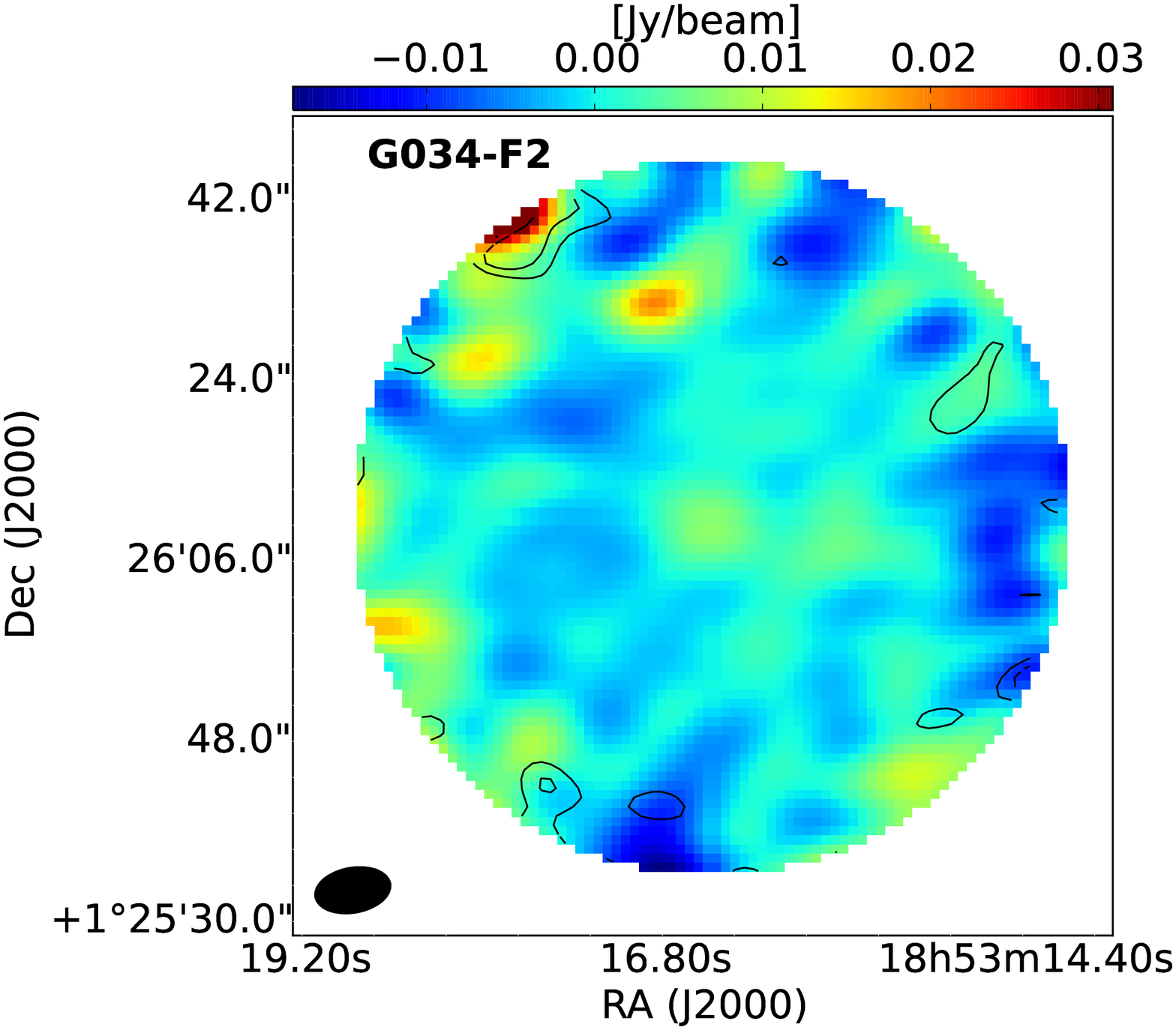}
 \includegraphics[width=0.7\columnwidth]{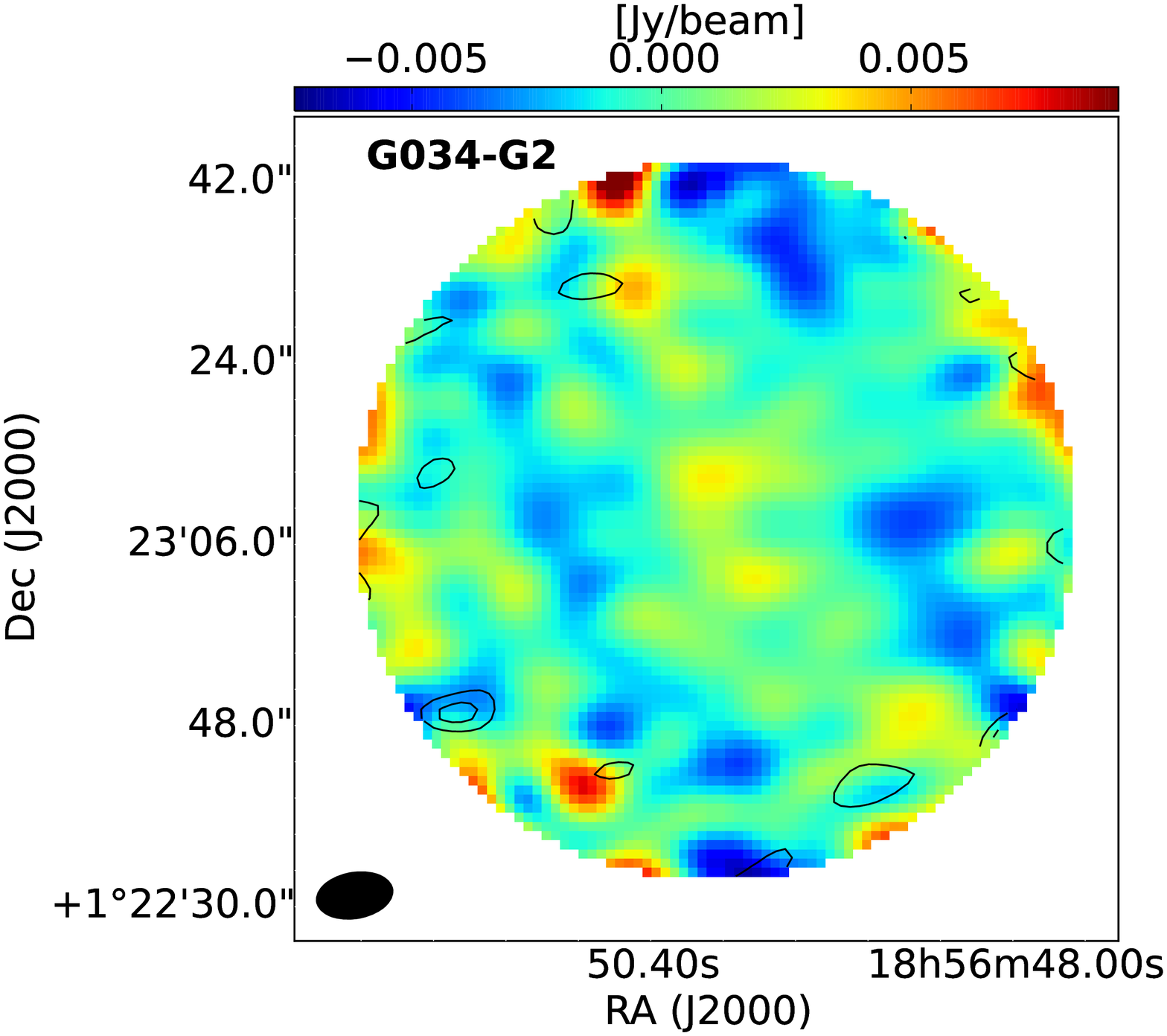}
  \includegraphics[width=0.7\columnwidth]{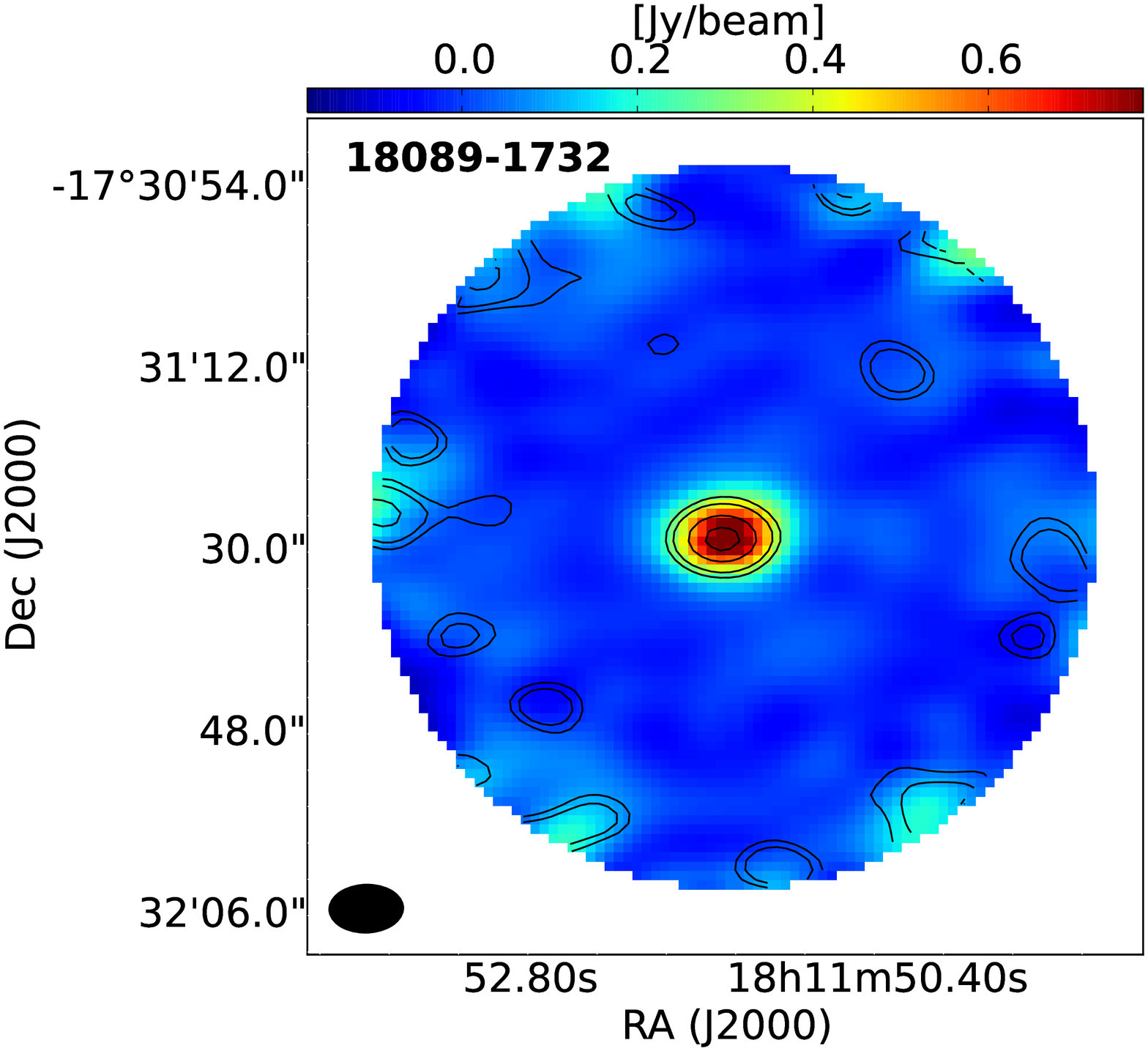}
  \includegraphics[width=0.7\columnwidth]{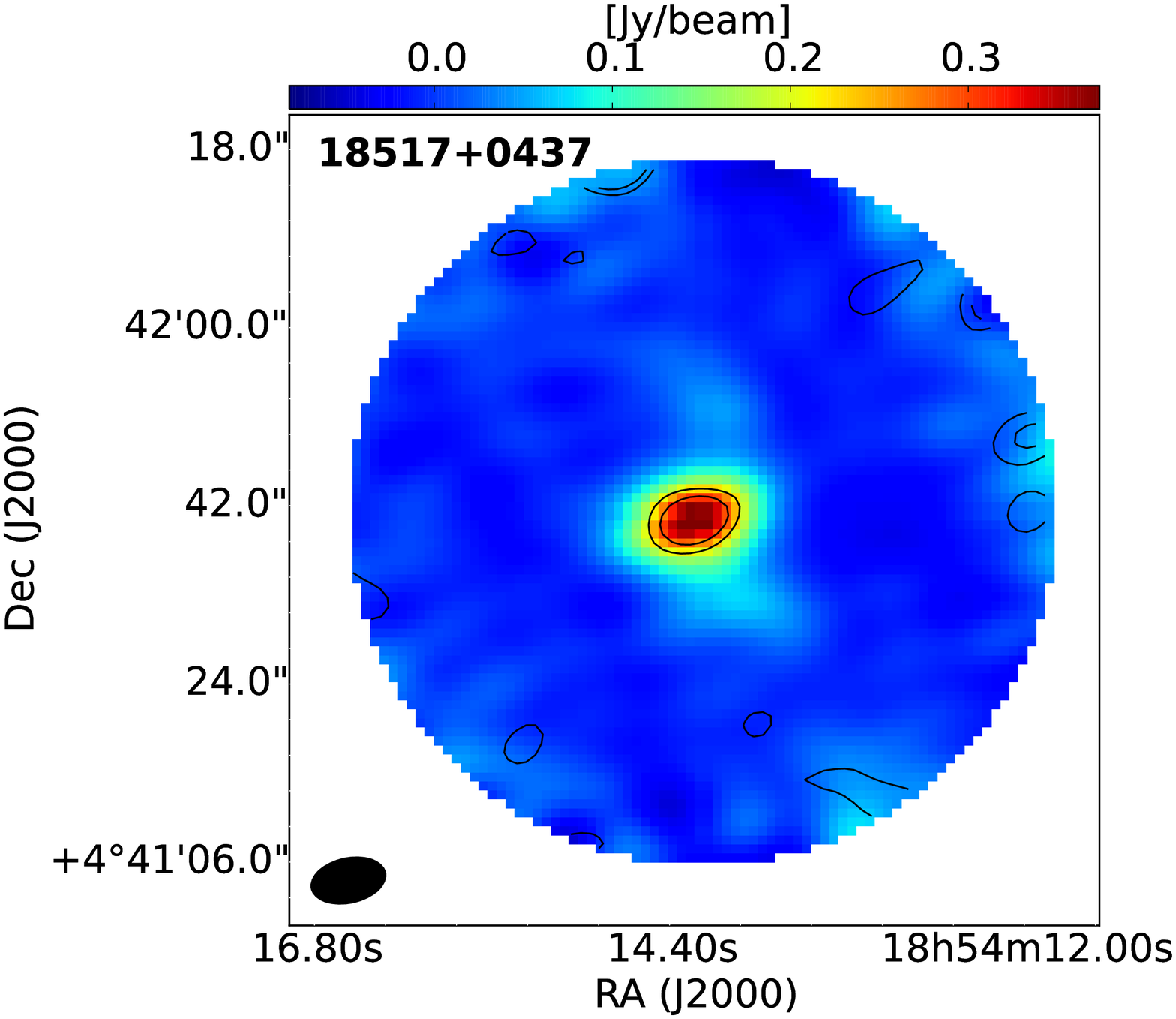}
  \includegraphics[width=0.7\columnwidth]{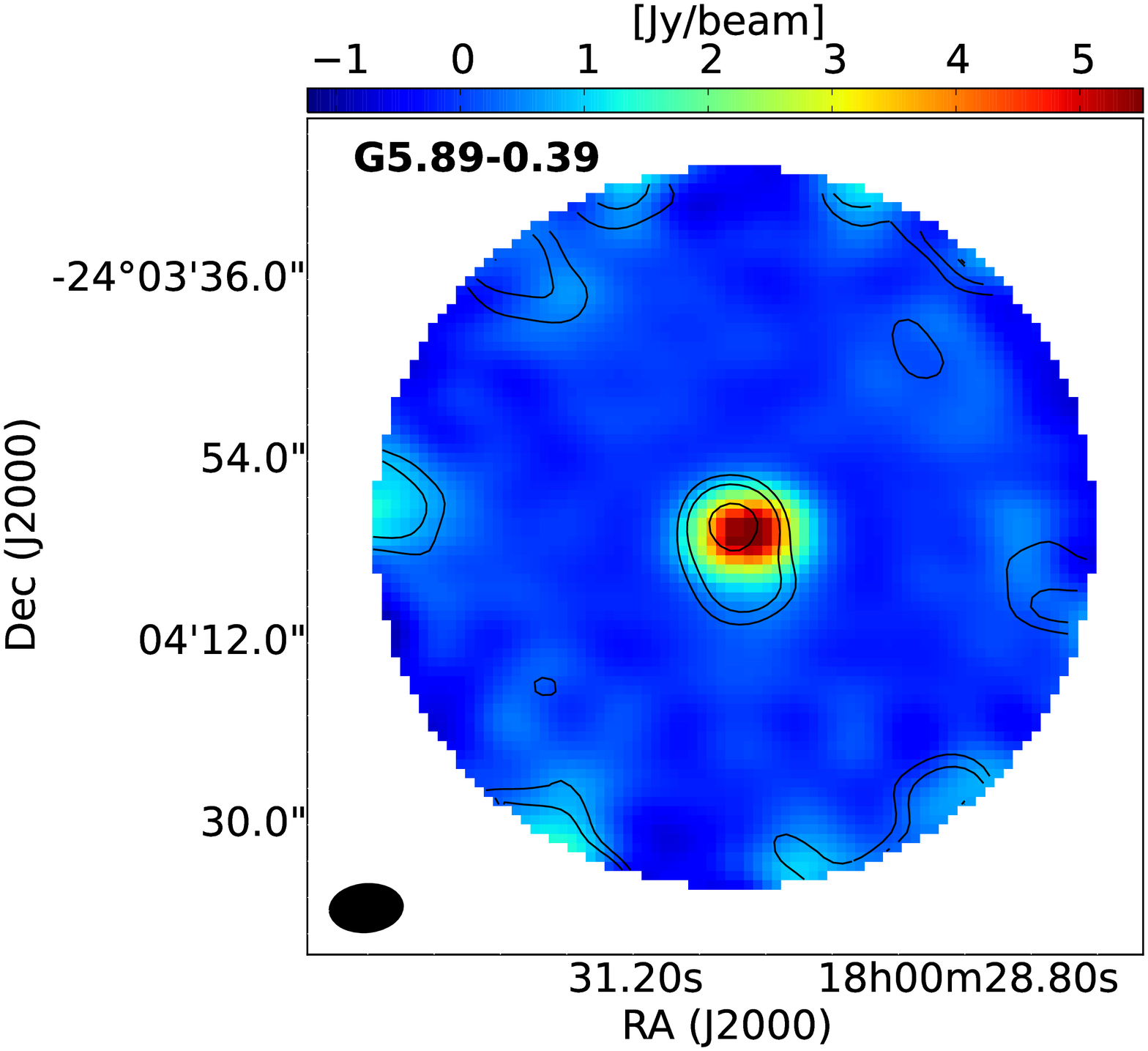}
  \includegraphics[width=0.7\columnwidth]{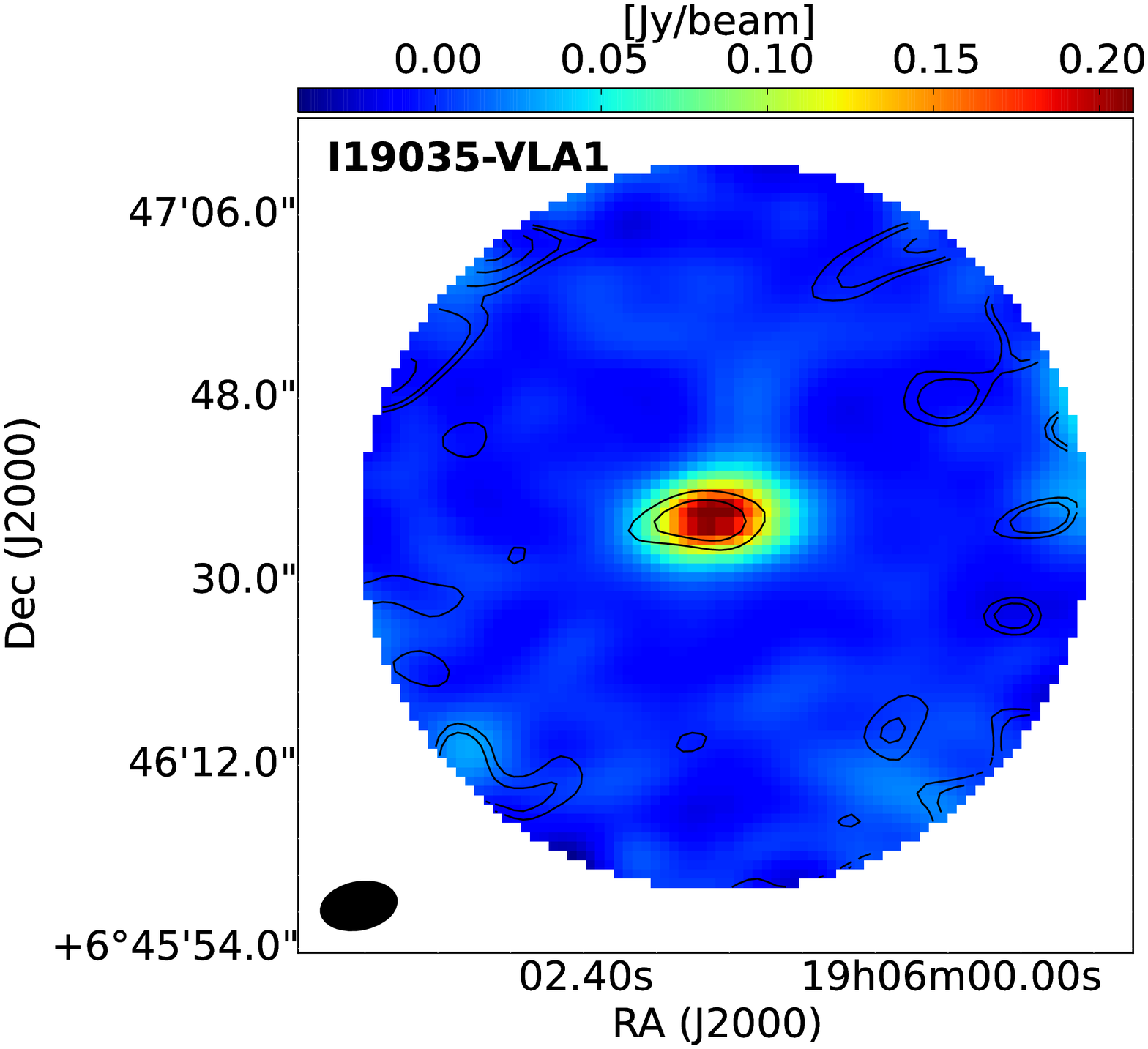}
  \includegraphics[width=0.7\columnwidth]{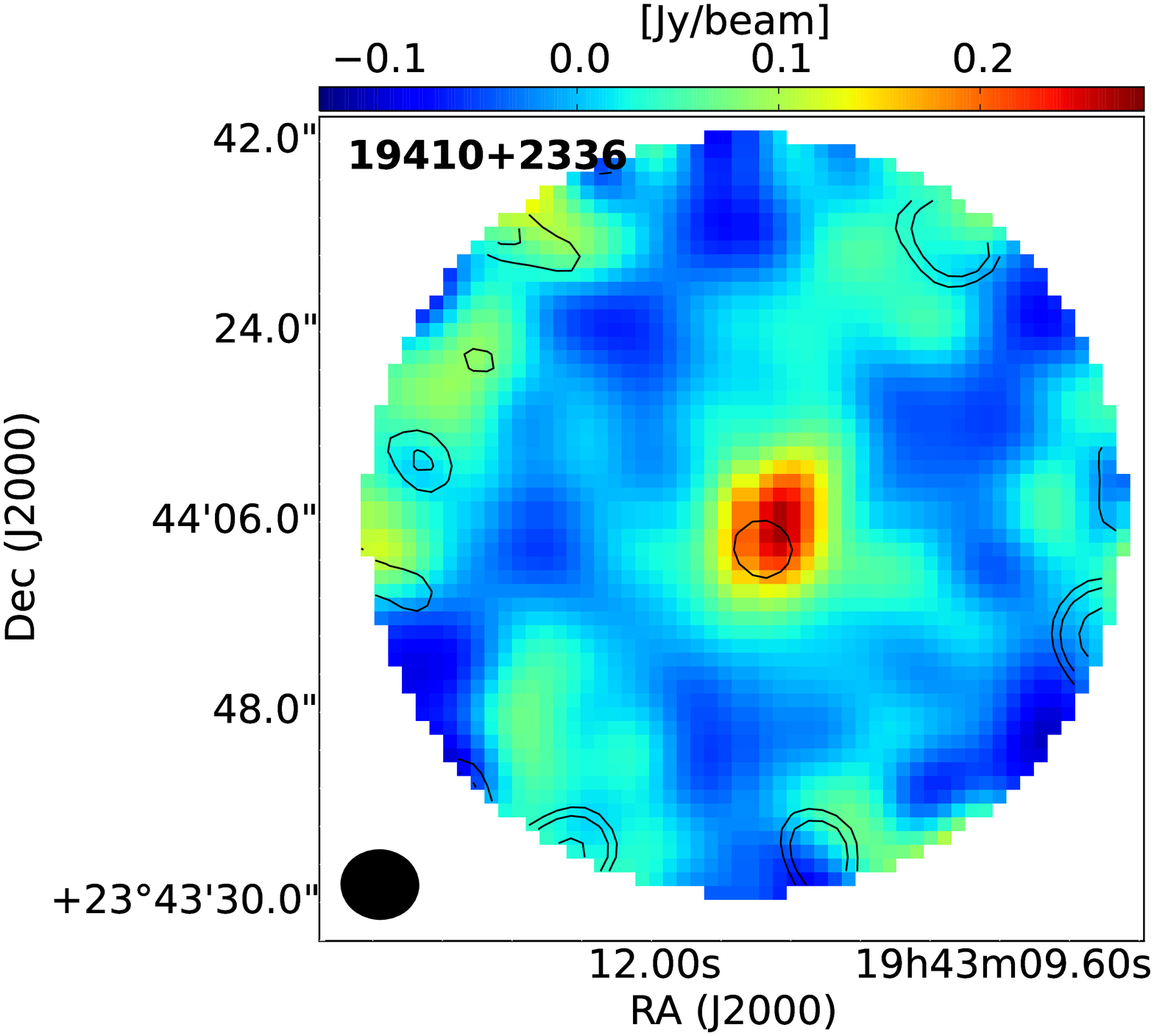}
   \caption{211\,GHz continuum maps toward G028-C1, G034-F2, G034-G2, 18089-1732, 18517+0437, G5.89-0.39, I19035-VLA1 and 19410+2336 (sorted by evolutionary stage). The black contours indicate the 3, 5, 10 and 15\,$\sigma$ levels of H$_2^{13}$CO 3$_{0,3}-2_{0,2}$ integrated intensity, where $\sigma$ is 0.1\,Jy/beam. No H$_2^{13}$CO 3$_{0,3}-2_{0,2}$ emission was detected toward G028-C1. Neither continuum nor H$_2^{13}$CO 3$_{0,3}-2_{0,2}$ emission was detected toward the cores in the G034 region. \label{figure:continuum_H213CO_maps}}
\end{figure*}

We have checked the H$_2^{13}$CO 3$_{0,3}-2_{0,2}$ and continuum emission of our sources in our Band\,6 ALMA data. The continuum emission was detected for 6 objects out of the 8 sources covered in our observations. We observed, but not detected compact continuum sources above 3$\sigma$ toward two sources in the youngest, HMSC evolutionary stage (G034-F2 and G034-G2) at our sensitivity and resolution. We note that \citet{Kong2017} observed these objects and detected 1.3\,mm continuum emission of $\sim$1-3\,mJy\,beam$^{-1}$ with a beam size of 1.5''$\times$1.0''. The H$_2^{13}$CO emission was not detected toward any of the HMSC sources. For the objects with detection, the H$_2^{13}$CO emission is associated with the same region as the continuum emission. Figure\,\ref{figure:continuum_H213CO_maps} shows the continuum maps at 213\,GHz and the H$_2^{13}$CO contours overplotted. We derived the effective radius of the sources based on the 2D Gaussians fitted to the continuum maps. The derived source sizes are between 3'' and 5.8''. We note that in \citet{Zahorecz2017} we used 6'' as the source size for AFGL5142, IRAS05358+3543 and G5.89-0.39. In our actual calculations, we updated the size of G5.89-0.39 based on its ALMA continuum emission. AFGL5142 and IRAS05358+3543 have no ALMA data and G028-C1, G034-F2 and G034-G2 have no H$_2^{13}$CO detection and / or continuum detection. We kept the assumption of 6'' source size for these five objects in our calculations.

\subsection{Molecular column densities and column density ratios}

To estimate the molecular column densities from the APEX and IRAM spectra of H$_2$CO, H$_2^{13}$CO, HDCO and D$_2$CO, we used the MADCUBA software that assumes LTE conditions \citep[][]{Martin2019}. We have used the SLIM (Spectral Line Identification and Modeling) tool to produce synthetic spectra of the observed molecules. When generating the synthetic spectra the beam size is taken into account by the software to return the appropriate line intensities. The MADCUBA-AUTOFIT tool was used to derive the physical conditions of the gas by comparing the observed spectra with the LTE synthetic spectra taking into account all transitions considered. The autofit tool uses the Levenberg-Marquardt algorithm to provide the best nonlinear least-squared fit \citep[see details in][]{Martin2019}. We adopted a source size of $3.\!^{\prime\prime}0$ - $6.\!^{\prime\prime}0$ according to the continuum sizes measured with ALMA and following \citet{Zahorecz2017} and \citet{Fontani2015} for sources without continuum detection. We derived the excitation temperature of the gas, T$_\mathrm{ex}$, using the four APEX HDCO transitions (Table~\ref{table:observed_transitions}). The derived average temperature is $\sim$32\,K with a standard deviation of 17\,K. We used these temperatures to fit the H$_2^{13}$CO, HDCO and D$_2$CO transitions. 
For the brightest sources, we could not fit all the detected H$_2$CO transitions using these low T$_\mathrm{ex}$ temperature. In this case, we used the H$_2$CO lines independently to derive their excitation temperature. For the youngest sources, the HMSCs, no HDCO emission was detected, therefore we derived the excitation temperature based on the H$_2$CO lines. We used the 3$\sigma$ noise levels to provide the upper limits to the intensity of the non-detected lines.

We have estimated the H$_2$CO column densities from the H$_2^{13}$CO column density using the $^{12}$C/$^{13}$C ratio (see in Table \ref{table:basic_data}) derived by \citet{Milam2005} as a function of Galactocentric distance to minimize the optical depth's effect.
In Table \ref{table:madcubaij_results}, we report the derived excitation temperatures and column densities of H$_2$CO, H$_2^{13}$CO, HDCO and D$_2$CO. 

In Table \ref{table:deuteration}, we report the calculated column density ratios of HDCO/H$_2$CO, D$_2$CO/H$_2$CO and D$_2$CO/HDCO. The column density ratio HDCO/H$_2$CO corresponds to the D$_\mathrm{frac}$(H$_2$CO) value. 
The presented column density ratios are source-averaged values. We note that the different source distance can in principle introduce biases in the comparison, because the linear scale observed is not the same. However, in our study we see from the interferometric images that the emission of the lines is more compact than the beam, and hence the emission arising from the (different) envelope region should not influence the derived source-averaged ratios in a significant way.
The D$_2$CO/HDCO column density ratios show no clear trend since they have a large scatter around the median value of 0.21 (standard deviation of 2.16). 
The N(HDCO)/N(H$_2$CO) and N(D$_2$CO)/N(H$_2$CO) ratios show a similar trend. The deuteration fraction of singly deuterated formaldehyde shows higher values, around 0.1 at the HMSC-HMPO stage and it decreases to 0.01 at the UC~\ion{H}{II} phase. The deuteration fraction of the doubly deuterated formaldehyde reaches 0.03 at the earliest stages and drops down to 0.003 at the most evolved phase.

\begin{table*}[!ht]							
\centering							
\begin{tabular}{l c c c c c}							\hline							
Source & Stage & \underline{N(HDCO)} & \underline{N(D$_2$CO)} & \underline{N(D$_2$CO)}	& F\\
& & N(H$_2$CO) & N(HDCO) & N(H$_2$CO) &  \\
\hline
G028-C1	& HMSC &	$<$0.297	&	$<$1.223	&	$<$0.364& -	\\
G034-F2	& HMSC & 	$<$0.513	&	$<$0.794	&	$<$0.408 & - \\
G034-F2	& HMSC &	$<$0.520	&	$<$0.794	&	$<$0.413 & - \\
G034-G2	& HMSC & 	$<$0.339	&	$<$0.794	&	$<$0.269 & - \\
G034-G2	& HMSC &	$<$0.688	&	$<$0.794	&	$<$0.546 & - \\
AFGL5142 & HMSC/HMPO	&	0.162$\pm$0.035$^*$	&	0.153$\pm$0.035	&	0.025$\pm$0.003$^*$	& 1.05$\pm$0.34 \\
AFGL5142 & HMSC/HMPO	&	0.105$\pm$0.017$^*$	&	0.268$\pm$0.051	&	0.028$\pm$0.004$^*$	& 0.39$\pm$0.11 \\
IRAS05358+3543 & HMSC/HMPO	&	0.122$\pm$0.018$^*$	&	0.180$\pm$0.031	&	0.022$\pm$0.003$^*$	& 0.68$\pm$0.17 \\
18089-1732 & HMPO	&	0.018$\pm$0.004$^*$	&	$<$0.055	&	$<$0.001$^*$ & $>$0.32	\\
18089-1732	& HMPO &	0.030$\pm$0.013$^*$	&	$<$0.066	&	$<$0.002$^*$ & $>$0.45\\
18517+0437	& HMPO &	0.304$\pm$0.144$^*$	&	0.097$\pm$0.031	&	0.029$\pm$0.016$^*$	& 3.19$\pm$2.77  \\
18517+0437 & HMPO &	0.001$\pm$0.001$^*$	&	5.397$\pm$2.540	&	0.005$\pm$0.003$^*$	& 0.0002$\pm$0.0003\\
G75-core & HMPO &	0.053$\pm$0.013$^*$	&	$<$0.103	&	$<$0.005$^*$	& $>$0.56\\
G5.89-0.39 & UC~\ion{H}{II} &	0.004$\pm$0.001$^*$	&	$<$0.080	&	$<$0.001$^*$	& $>$0.02\\
G5.89-0.39 & UC~\ion{H}{II} &	0.008$\pm$0.001$^*$	&	$<$0.0282	&	$<$0.001$^*$	& $>$0.06\\
I19035-VLA1	& UC~\ion{H}{II} &	$<$0.021$^*$	&	$<$0.759	&	$<$0.016$^*$ & - \\
19410+2336	& UC~\ion{H}{II} &	0.021$\pm$0.006$^*$	&	0.206$\pm$0.072	&	0.004$\pm$0.001$^*$	& 0.11$\pm$0.05 \\
\hline							
\end{tabular}							
\caption{Calculated column density ratios for the observed species. We note that $^*$ sign at the numbers indicate that the values calculated based on the H$_2^{13}$CO line assuming a $^{12}$C/$^{13}$C ratio according to \citet{Milam2005}. Each velocity component is shown as a separate row. \label{table:deuteration}}		
\end{table*}

\section{Discussion}
In this section we compare our results with those inferred from previous observational studies (Section \ref{section:comparison_previous_obs}). We also put our results in context with theoretical studies of formaldehyde deuteration via grain-surface and gas-phase processes (Section \ref{section:grain_surface_deuteration}).

\subsection{Comparison with previous observational studies \label{section:comparison_previous_obs}}
\citet{Fontani2011} presented the kinetic temperatures based on ammonia data. Nine sources from our sample have available ammonia temperatures with values lying between $\sim$20-40\,K, except for the G75-core, which has a higher kinetic temperature of 96\,K. The HDCO-based temperature estimates are in good agreement with the ammonia temperatures for most of the sources: they are in the 15-40\,K range. G5.89-0.39 and I19035-VLA1 show higher HDCO excitation temperatures, $\sim$60-70\,K. For the youngest HMSC sources, we only detected H$_2$CO. The excitation temperatures fall into the 20-30\,K range.
For some of the sources, the HDCO derived temperatures were too low to fit the line intensity ratios of the observed H$_2$CO transitions. Also in these cases, we derived the H$_2$CO excitation temperature, finding values a factor of 2-3 higher than the HDCO values, in the range of $\sim$70-130\,K. 

\citet{Fontani2015} have proposed that in high-mass star-forming regions the different trend for the deuteration pattern of different molecules such as N$_2$H$^+$, HNC, NH$_3$ and CH$_3$OH as a function of evolutionary stage, is likely due to the way deuteration occurs for the different species: for N$_2$H$^+$, detuteration occurs in the gas phase, while for CH$_3$OH it occurs on grain surfaces; for NH$_3$, deuteration likely occurs via a mixture of these two processes. In the following, we discuss the likelihood of each of these two chemical routes to drive the formation of the deuterated forms of H$_2$CO during the massive star formation process.

\subsection{Grain-surface deuteration of H$_2$CO \label{section:grain_surface_deuteration}}

Traditionally, it has been assumed that the deuterated forms of H$_2$CO are a product of grain-surface chemistry. Indeed, \citet{Taquet2012} showed that the gas-phase formation time-scales of D$_2$CO are longer than the depletion time-scales in the pre-stellar phase and, therefore, one would expect D$_2$CO to be a product of material processed on solid ices. Consequently, the N(D$_2$CO)/N(H$_2$CO) ratio should show an evolutionary trend similar to that derived for $D_\mathrm{frac}$(CH$_3$OH), which peaks at the HMPO phase \citep{Fontani2015}. If HDCO showed a similar behaviour to D$_2$CO, we could conclude that HDCO also forms mostly on dust grains. This hypothesis would be supported by the fact that HDCO is detected in hot cores and hot corinos while N$_2$D$^+$ is not \citep[see e.g.][]{Fuente2005}.

\begin{figure}[!ht]
  \centering
   \includegraphics[width=1.0\columnwidth]{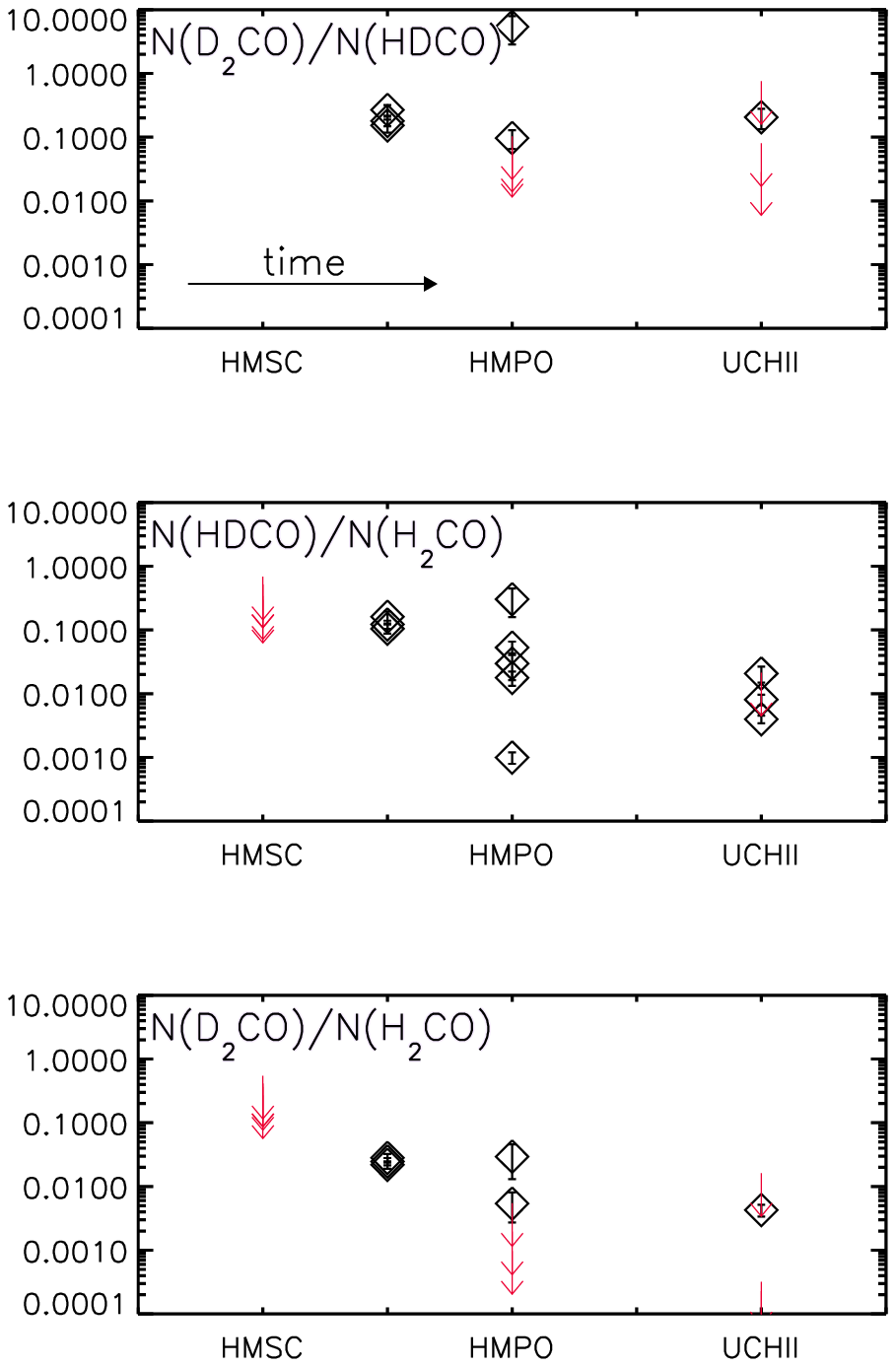}
  \caption{Comparison of observed column density ratios of H$_2$CO, HDCO and D$_2$CO. Upper limits are shown with red symbols. The arrow in the top panel indicates the increasing time from the HMSC phase towards the UC~\ion{H}{II} phase. \label{figure:dfrac_comparison}}
\end{figure}

Fig. \ref{figure:dfrac_comparison} shows that the column density ratios of N(HDCO)/N(H$_2$CO) and N(D$_2$CO)/N(H$_2$CO) decrease from the HMSC-HMPO phase to the UC~\ion{H}{II} phase by factors of $\geq$10. We note that for the two sources in the HMSC/HMPO transitional category (AFGL5142 and IRAS05358+3543), we cannot disentangle the contribution of the HMSC cores from the HMPO ones. However, from our observations it is clear that they show higher column density ratios than those found for the UC~\ion{H}{II} regions, confirming the decrease of the N(HDCO)/N(H$_2$CO) and N(D$_2$CO)/N(H$_2$CO) trend. Based on the available upper limits for the HMSC cores, we cannot exclude the possibility that the D$_2$CO/H$_2$CO column density ratio peaks in the early phases. However, the upper limits obtained for the HMSC cores suggest a lower or, at most, equal HDCO/H$_2$CO column density ratios than those of the HMPO objects. This behaviour is similar to the trend observed for $D_\mathrm{frac}$(CH$_3$OH) but with a larger scatter. This suggests that grain surface chemistry may be the main formation mechanism for HDCO and D$_2$CO.

\citet{Rodgers2002} proposed a grain surface formation scenario for these molecules in which H$_2$CO and its deuterated forms represent intermediate steps in the formation of CH$_3$OH and deuterated-CH$_3$OH via hydrogenation and D-addition reactions. They defined the F parameter (see Equation 1 below) to investigate the extent by which surface chemistry is responsible for the production of HDCO and D$_2$CO. \citet{Rodgers2002} found that gas-phase chemistry should produces F values between 1.6 and 2.3, and grain surface chemistry should result in F values $\geq$ 1. 
Previous observations, however, have reported F values lower than 1, which are not consistent with the cases presented above. Alternatively, low F values could be obtained through a grain surface chemistry over a long time period \citep{Rodgers2002}, because longer time-scales allow the conversion of HDCO into D$_2$CO on the surface of dust grains.
We derived the F parameter based on our calculated HDCO/H$_2$CO and D$_2$CO/H$_2$CO ratios using the definition of \citet{Rodgers2002}:
\begin{equation}
F=\frac{[\mathrm{HDCO}/\mathrm{H}_2\mathrm{CO}]^2}{[\mathrm{D}_2\mathrm{CO}/\mathrm{H_2CO}]}.
\label{Fpar}
\end{equation}

\noindent
In our sample sources with D$_2$CO detections, F lies between 0.1 and 1.1 (except 18517+0437, where the two fitted components show F values $\sim$3 and $\sim$0.001). For the sources with only upper limits for the detection of D$_2$CO, the lower limits for the F values are within 0.1 and 0.6.
These values cannot be explained by gas phase chemistry, but are more consistent with grain surface chemistry models.

Table~\ref{table:deuteration} also shows that the D$_2$CO/HDCO ratio remains constant at $\sim$0.2, within the uncertainties, regardless of the source or evolutionary stage (except for 18517+0437). As for the low-mass regime, the ratio HDCO/D$_2$CO lies well below the statistical value given as D-species/D$_2$-species=4$\times$(D-species/H-species)$^{-1}$, consistent with the grain surface formation scenario \citep[see][]{Ceccarelli2014}.

\citet{Ceccarelli2014} calculate the column density ratios for HDCO/H$_2$CO and D$_2$CO/H$_2$CO for a sample of low-mass star-forming cores (see their Figure 5). The derived ratios lie close to 1 for those objects. The average derived $\frac{\mathrm{HDCO}/\mathrm{H_2CO}}{\mathrm{D_2CO}/\mathrm{H_2CO}}$ ratio is $\sim$5.3 with a standard deviation of 3.4 (1.1 if we exclude the two velocity components of the 18517+0437 region, showing a large difference from the other sources) for our high-mass star-forming regions. If we compare these findings with the results of \citet{Ceccarelli2014}, it is possible that the higher values measured toward our sample is the result of the shorter time-scales available in the high-mass regime for the formation of D$_2$CO in the ices \citep{Taquet2012, Rodgers2002}. In fact, the HDCO/H$_2$CO ratios are factors $\sim$10 or more larger than $D_\mathrm{frac}$(CH$_3$OH) in high-mass star-forming regions, which is also consistent with the shorter time-scales available for deuteration in high-mass star formation with respect to the low-mass case, and with the formation of these species via hydrogenation and D-addition reactions on grain surfaces \citep{Turner1990, Rodgers2002}.

\citet{Turner1990} reported the first interstellar detection of a multiply deuterated molecule, D$_2$CO in the Orion KL ridge component with a D$_2$CO/H$_2$CO column density ratio of 0.021. The calculated HDCO/H$_2$CO column density ratio was 0.14. These ratios could not be reproduced by ion-molecule gas phase reactions, but they can be fitted by the model of grain surface reactions of \citet{Tielens1983}. Additional studies toward dark clouds TMC-1 and L183 were performed by \citet{Turner2001} and they reported HDCO / H$_2$CO column density ratios of 0.059 and 0.068, respectively.

In addition, \citet{Fontani2014} successfully detected HDCO towards the protostellar bow-shock L1157-B1 where material from icy mantles is thought to be freshly released due to mantle evaporation. They have found a column density ratio of HDCO/H$_2$CO value of 0.1, similar to the values found in our study. \citet{Bianchi2017} studied the deuteration in Sun-like protostars. Their Figure 9 shows measured column density ratios of HDCO/H$_2$CO as a function of the bolometric luminosity for both low-mass and high-mass star-forming regions from the protostellar phase to the Class I phase. They have also found that the average column density ratio measured in Class 0 sources for HDCO is $\sim$0.12. However, D$_2$CO shows an increase going from pre-stellar cores, with an average value of N(D$_2$CO/H$_2$CO)$\sim$0.045, to Class 0 sources ($\sim$0.15) and then a strong decrease in the Class I protostar phase \citep[see the case of SVS13-A;][]{Bianchi2017}.
Our results indicate a similar, but not identical, evolutionary behaviour since no clear detection of deuterated formaldehyde is available for the HMSC sources.

As part of the ALMA-PILS survey, \citet{Persson2018} and \citet{Manigand2020} have reported the deuterium fraction of formaldehyde for IRAS 16293-2422 A and B. Their relative evolutionary stages are under debate, although recent studies suggest that the B component is at an early evolutionary stage \citep{vanderWiel2019}. Formaldehyde abundance is lower towards the A component with a HDCO/H$_2$CO column density ratio of around 0.049 and 0.065 for the A and B components, respectively. They agree well with our results for the HMPO sources. \citet{Persson2018} and \citet{Manigand2020} have also found that the doubly-deuterated isotopologue D$_2$CO is significantly enhanced compared to the singly-deuterated isotopologue toward the B component. The calculated D$_2$CO/H$_2$CO column density ratios are 0.041 and 0.006, similar to the values obtained for our HMPO sources.
For the B source, the column density ratio of D$_2$CO/HDCO is slightly larger than the column density ratio HDCO/H$_2$CO, which is consistent with formaldehyde forming in the ice as soon as CO is frozen onto grains.

\begin{figure*}[!ht]
  \centering
\includegraphics[width=2.0\columnwidth]{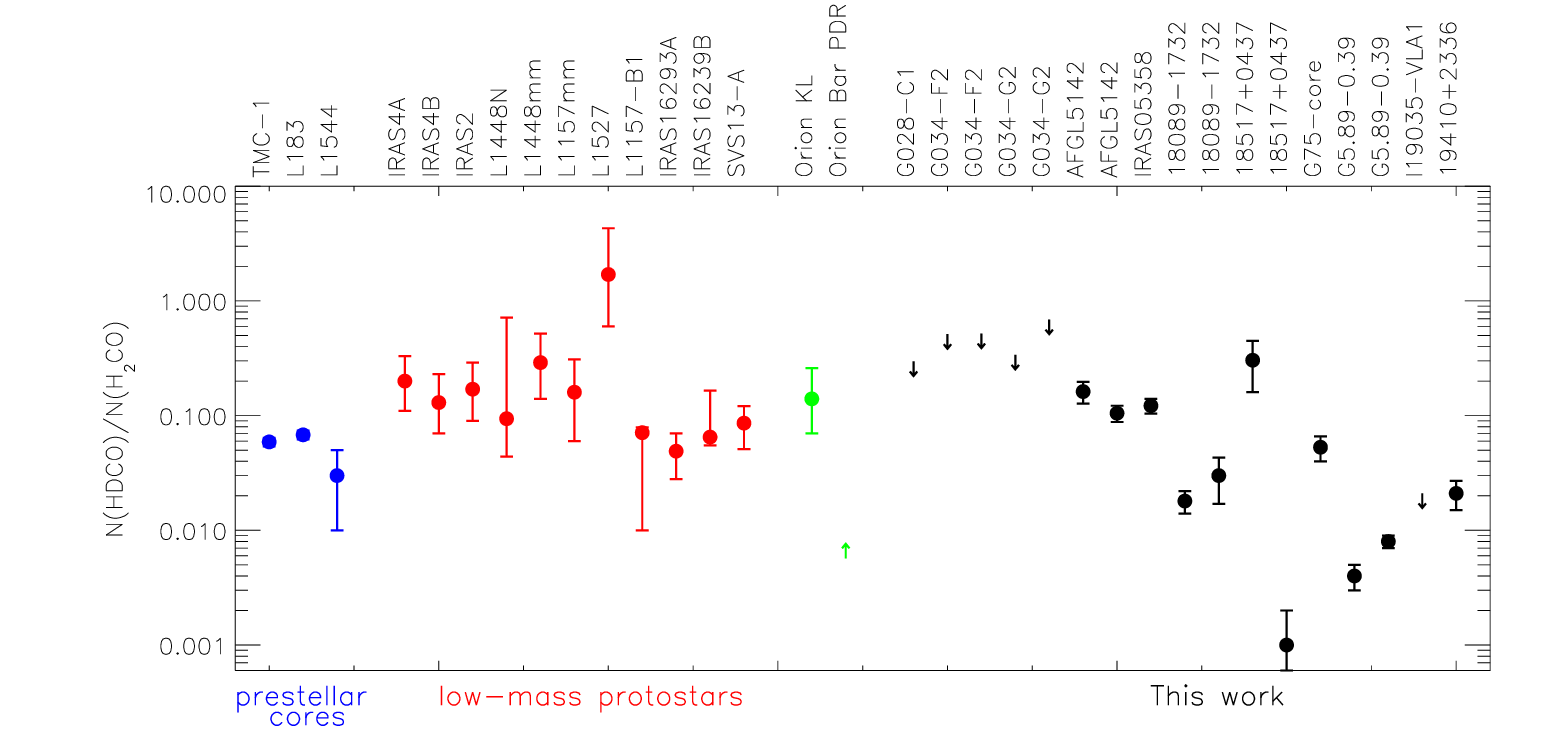}   
\includegraphics[width=2.0\columnwidth]{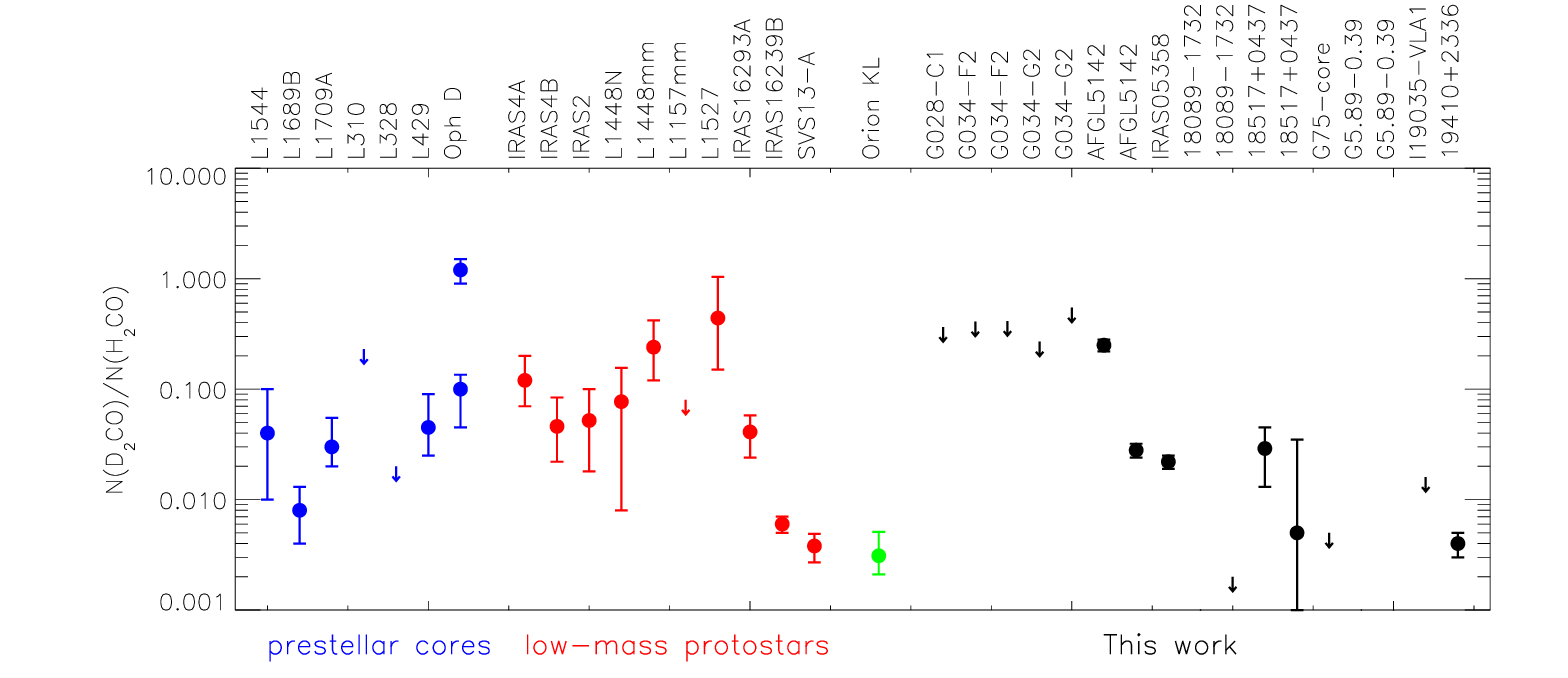}
  \caption{Comparison of the observed N(D$_2$CO)/N(H$_2$CO) (top panel) and N(D$_2$CO)/N(H$_2$CO) (bottom panel) column density ratios for our sources and for prestellar cores \citep{Turner2001, Bacmann2003, Chacon2019}, low-mass protostars \citep{Parise2006, Ceccarelli2014, Bianchi2017, Fontani2014, Manigand2020}, and sources in the Orion region \citep{Turner1990, Parise2009} region. \label{figure:compare_literature}}
\end{figure*}

Figure \ref{figure:compare_literature} shows the comparison of the observed H$_2$CO, HDCO and D$_2$CO column density ratios in different environments based on the works mentioned in this section.

\subsection{Gas-phase deuteration of H$_2$CO}

Alternatively to the grain-surface formation of HDCO and D$_2$CO, the deuterated forms of H$_2$CO could be produced by gas-phase chemistry. \citet{Roberts2007} and \citet{Roueff2007} proposed that deuterium fractionation of H$_2$CO may occur via the fractionation of CH$_3^+$ (CH$_2$D$^+$ $\rightarrow$ CH$_4$D$^+$ $\rightarrow$ CH$_2$D $\rightarrow$ HDCO) or of H$_2$CO directly via the reaction with H$_2$D$^+$. In particular, in \cite{Bergman2011}'s gas-phase steady-state models based on the reaction network of \cite{Roueff2007}, these authors also predict a decrease of the deuteration of H$_2$CO from the HMSC/HMPO stage to the UC~\ion{H}{II} region phase as the temperature exceeds 40\,K. The predicted values for the D/H ratio inferred from HDCO/H$_2$CO and D$_2$CO/HDCO ratios are indeed consistent with our observations (see their Figure 12). Indeed, while the HDCO/H$_2$CO ratio gives $\sim$0.1 for T$\sim$10-30 K and $\leq$0.04 for T$\geq$40 K, the D$_2$CO/H$_2$CO ratio remains relatively flat at $\sim$0.2 for T$\sim$10-40 K, as observed in our sample. The only difference between our observations and the predictions by \cite{Bergman2011} is that, while in our observations the HDCO/H$_2$CO ratio drops by more than one order of magnitude from the HMPO to the UC~\ion{H}{II} region stage, the chemical modelling predicts a smaller drop by only factors 3-4. \citet{Parise2009} found high deuterium fractionation toward the Orion Bar PDR region. The calculated deuterium fractions are consistent with predictions of pure gas-phase chemistry models at warm temperatures. They have found an N(HDCO)/N(H$_2$CO) ratio of 0.0057. This value is in the range found for the UC~\ion{H}{II} regions in our sample.

In any case, to distinguish between the grain-surface from the gas-phase scenarios we would need to obtain stringent upper limits to the HDCO/H$_2$CO ratios at the HMSC stage. From our observations, these limits can get as low as $\leq$0.1 when considering the rms measured over the whole linewidth of the lines, which is still consistent with the gas-phase results from \cite{Bergman2011}'s models (grain-surface deuteration is expected to give much lower upper limits as a consequence of freeze-out in the cold conditions of HMSC). Therefore, although our results suggest that grain surface deuteration is the most likely mechanism for the formation of HDCO and D$_2$CO in star-forming regions, higher sensitivity observations are needed to provide more stringent constraints to the column density ratio of HDCO/D$_2$CO during the HMSC phase.

\section{Summary}
We have observed 213\,GHz continuum, H$_2^{13}$CO, HDCO and D$_2$CO emission with APEX and ALMA toward a sample of high-mass star-forming regions at different evolutionary stages. HDCO transitions were detected for all of them except the three youngest sources in the HMSC stage, while the D$_2$CO line was detected only for four sources from the HMSC-HMPO stage to the UC~\ion{H}{II} stage. The H$_2^{13}$CO emitting regions are similar to the continuum source sizes, they are in the range of 3'' - 5.8'' corresponding to 0.02-0.07\,pc at the distance of the investigated regions. The column density ratio of HDCO / H$_2$CO shows the highest values of $\sim$0.13 in the earlier evolutionary stages (HMSC-HMPO) and then decreases by an order of magnitude toward the UC~\ion{H}{II} phase to $\sim$0.01. This study confirms the idea that H$_2$CO, and its deuterated species, form mostly on grain surfaces although some gas-phase contribution is expected at the warm HMPO stage. The calculated upper limits for the HMSC sources are high, around 0.5, so we cannot constrain the trend between HMSC and HMPO phases. More sensitive follow-up observations are needed for the HMSC sources to better constrain the upper limits. The observed column density ratios are in good agreement with the values previously reported in the literature for low-mass prestellar cores, protostellar objects and PDR regions.

\begin{acknowledgements}
This paper makes use of the following ALMA data: ADS/JAO.ALMA\#2017.1.01157.S. ALMA is a partnership of ESO (representing its member states), NSF (USA) and NINS (Japan), together with NRC (Canada), MOST and ASIAA (Taiwan), and KASI (Republic of Korea), in cooperation with the Republic of Chile. The Joint ALMA Observatory is operated by ESO, AUI/NRAO and NAOJ. 
This research was partially supported by the NAOJ ALMA Scientific Research Grant Number 2016-03B, the Italian Ministry of Education, Universities and Research through the grant Progetti Premiali 2012 iALMA (CUP C52I13000140001), the Deutsche Forschungsgemeinschaft (DFG, German Research Foundation) - Ref no. FOR 2634/1 TE 1024/1-1, the DFG cluster of excellence ORIGINS (www.origins-cluster.de), the EU Horizon2020 research and innovation programme, Marie Sklodowska-Curie grant agreement 823823 (Dustbusters RISE project), and the European Research Council (ERC) via the ERC Synergy Grant ECOGAL (grant 855130). Ke Wang acknowledges support by the National Key Research and Development Program of China (2017YFA0402702), the National Science Foundation of China (11721303), and the starting grant at the Kavli Institute for Astronomy and Astrophysics, Peking University (7101502016). I.J.-S. has received partial support from the Spanish FEDER (project number ESP2017-86582-C4-1-R) and the State Research Agency (AEI; project number PID2019-105552RB-C41).

\end{acknowledgements}

\bibliographystyle{aa}
\bibliography{references}

\appendix
\section{Observed H$_2$CO, HDCO, and D$_2$CO lines}
\label{app:lines}

In Table~\ref{table:madcubaij_results} and we report the parameters derived fitting single or multiple velocity components to the detected lines with MADCUBA.

\onecolumn
\begin{center}
\tablefirsthead{
\hline
Source	& Size &  Species	& N & T$_\mathrm{ex}$ & v$_\mathrm{LSR}$ & T$_\mathrm{NH_3}$\\
& [''] ([pc]) & & [10$^{13}$ cm$^{-2}$] &  [K] & [km/s] & [K]\\
\hline										
}
\tablehead{
\hline
Source	& Size & Species	& N & T$_\mathrm{ex}$ & v$_\mathrm{LSR}$ & T$_\mathrm{NH_3}$\\
& [''] ([pc]) & & [10$^{13}$ cm$^{-2}$] &  [K] & [km/s] & [K]\\
\hline										
}
\tabletail{
\hline}
\tablelasttail{
\hline}
\bottomcaption{Fitted H$_2$CO, H$_2^{13}$CO, HDCO and D$_2$CO excitation temperatures and column densities. We note that to estimate the upper limits of the HDCO and / or D$_2$CO column densities for G034-F2, G034-G2, 18089-1732, and G5.89-0.39, we used one component only and we used an average excitation temperature based on the successful fits of H$_2$CO / HDCO. No error bar is provided for T$_\mathrm{ex}$ in these cases. Source velocities and sizes are also indicated. The last column shows the kinetic temperatures adopted from \citet{Fontani2011}. \label{table:madcubaij_results}}
\begin{supertabular}{l r l c c c l}
\multicolumn{7}{c}{HMSC} \\	
\hline
G028-C1	& 6.0 (0.15) &	H$_2$CO	&	2.72$\pm$0.34	&	30.2$\pm$4.9	&	80.1$\pm$0.1 & 17	\\
	& 6.0 (0.15) &  HDCO	&	$<$0.81	&	30.2$\pm$4.9	&		& 17\\
	& 6.0 (0.15) & D$_2$CO	&	$<$0.99	&	30.2$\pm$4.9	&		& 17\\
G034-F2	& 6.0 (0.11) &  H$_2$CO	&	1.23$\pm$0.30	&	25.2$\pm$8.2	&	58.9$\pm$0.5	& \\
	& 6.0 (0.11) &	&	1.21$\pm$0.31	&	31.2$\pm$10.8	&	56.8$\pm$0.5	& \\
	& 6.0 (0.11) &	HDCO	&	$<$0.63	&	30.5	&		& \\
	& 6.0 (0.11) & 	D$_2$CO	&	$<$0.50	&	30.5	&		& \\
G034-G2	& 6.0 (0.08) &	H$_2$CO	&	1.48$\pm$0.12	&	19.0$\pm$4.8	&	42.0$\pm$0.2	& \\
	& 6.0 (0.08) &		&	0.73$\pm$0.01	&	29.6$\pm$12.8	&	43.8$\pm$0.2	& \\
	& 6.0 (0.08) &	HDCO	&	$<$0.50	&	24	&		& \\
	& 6.0 (0.08) &	D$_2$CO	&	$<$0.40	&	24	&		& \\
\hline
\multicolumn{7}{c}{HMSC/HMPO} \\
\hline
AFGL5142 & 6.0 (0.05)	&	H$_2$CO	&	876.51$\pm$57.11	&	13.6$\pm$0.9	&	-1.7$\pm$0.1	& 25\\
	& 6.0 (0.05) &		&	401.61$\pm$14.18	&	22.4$\pm$6.4	&	-3.8$\pm$0.1	& 25\\
	& 6.0 (0.05) & 	H$_2^{13}$CO	&	2.03$\pm$0.14	&	13.6$\pm$0.9	&	-2.2$\pm$0.2	& 25\\
	& 6.0 (0.05) &		&	1.17$\pm$0.10	&	22.4$\pm$6.4	&	-3.8$\pm$0.2	& 25\\
	& 6.0 (0.05) & 	HDCO	&	18.79$\pm$3.82	&	13.6$\pm$0.9	&	-2.2$\pm$0.2	& 25\\
	& 6.0 (0.05) & 		&	7.03$\pm$0.99	&	22.4$\pm$6.4	&	-3.8$\pm$0.2	& 25\\
	& 6.0 (0.05) &	D$_2$CO	&	2.88$\pm$0.32	&	13.6$\pm$0.9	&	-2.2$\pm$0.2	& 25\\
	& 6.0 (0.05) &		&	1.89$\pm$2.42	&	22.4$\pm$6.4	&	-3.8$\pm$0.2	& 25\\
IRAS05358+3543 & 6.0 (0.05)	&	H$_2$CO	&	100.51$\pm$3.68	&	71.9$\pm$3.3	&	-16.2$\pm$0.1	& 35\\
	& 6.0 (0.05) &	H$_2^{13}$CO	&	1.96$\pm$0.14	&	27.2$\pm$6.0	&	-16.2$\pm$0.1	& 35\\
	& 6.0 (0.05) &	HDCO	&	13.69$\pm$1.75	&	27.2$\pm$6.0	&	-16.1$\pm$0.1	& 35\\
	& 6.0 (0.05) &	D$_2$CO	&	2.46$\pm$0.28	&	27.2$\pm$6.0	&	-16.4$\pm$0.2	& 35 \\
\hline
\multicolumn{7}{c}{HMPO} \\
\hline
18089-1732 & 3.6 (0.06)	&	H$_2$CO	&	360.46$\pm$7.49	&	84.7$\pm$1.5	&	35.0$\pm$0.5	& 38\\
	& 3.6 (0.06) &		&	184.09$\pm$5.28	&	80.8$\pm$2.4	&	32.3$\pm$0.5	& 38\\
	& 3.6 (0.06) &	H$_2^{13}$CO	&	8.91$\pm$0.65	&	18.7$\pm$11.8	&	34.2$\pm$0.2	& 38\\
	& 3.6 (0.06) &		&	4.42$\pm$0.49	&	22.5$\pm$17.4	&	31.9$\pm$0.2	& 38\\
	& 3.6 (0.06) & 	HDCO	&	7.28$\pm$1.76	&	18.7$\pm$11.8	&	34.2$\pm$0.2	& 38\\
	& 3.6 (0.06) & 		&	6.04$\pm$2.63	&	22.5$\pm$17.4	&	31.9$\pm$0.2	& 38 \\
	& 3.6 (0.06) &	D$_2$CO	&	$<$0.39	&	20	&		& 38\\
18517+0437 & 5.3 (0.07)	&	H$_2$CO	&	137.19$\pm$16.22	&	22.6$\pm$6.5	&	44.2$\pm$0.5	& \\
	& 5.3 (0.07) &		&	20.59$\pm$3.03	&	23.8$\pm$9.2	&	45.1$\pm$0.5	& \\
	& 5.3 (0.07) & 	H$_2^{13}$CO	&	1.03$\pm$0.48	&	22.6$\pm$6.5	&	44.0$\pm$0.1	& \\
	& 5.3 (0.07) &		&	2.66$\pm$0.48	&	23.8$\pm$9.2	&	44.6$\pm$0.2	& \\
	& 5.3 (0.07) &	HDCO	&	15.73$\pm$2.00	&	22.6$\pm$6.5	&	43.9$\pm$0.1	& \\
	& 5.3 (0.07) &		&	0.13$\pm$0.02	&	23.8$\pm$9.2	&	45.7$\pm$0.5	& \\
	& 5.3 (0.07) &	D$_2$CO	&	1.53$\pm$0.47	&	22.6$\pm$6.5	& 43.8$\pm$0.5	& \\
	& 5.3 (0.07) &		&	0.72$\pm$0.33	&	23.8$\pm$9.2	&	44.9$\pm0.5$	& \\
G75-core & 6.0 (0.11) 	&	H$_2$CO	&	152.4$\pm$4.79	&	34.9$\pm$12.3	&	0.5$\pm$0.1	& 96\\
	& 6.0 (0.11) &		&	16.79$\pm$2.18	&	34.9$\pm$12.3	&	-4.0$\pm$0.2	& 96\\
	& 6.0 (0.11) &	H$_2^{13}$CO	&	5.12$\pm$0.11	&	34.9$\pm$12.3	&	-0.1$\pm$0.1	& 96\\
	& 6.0 (0.11) &	HDCO	&	12.18$\pm$2.89	&	34.9$\pm$12.3	&	-0.1$\pm$0.2	& 96\\
	& 6.0 (0.11) &	D$_2$CO	&	$<$1.26	&	34.9$\pm$12.3	&		& 96\\
\hline
\multicolumn{7}{c}{UC~\ion{H}{II}} \\				
\hline
G5.89-0.39 & 3.0 (0.02)	&	H$_2$CO	&	8062$\pm$732	&	132.6$\pm$10.5	&	11.5$\pm$0.2	& \\
	& 3.0 (0.02) &		&	2667$\pm$567	&	86.0$\pm$7.6	&	8.4$\pm$0.2	& \\
	& 3.0 (0.02) &	H$_2^{13}$CO	&	130.30$\pm$12.32	&	70.0$\pm$7.5	&	11.6$\pm$0.2	& \\
	& 3.0 (0.02) &	HDCO	&	31.32$\pm$3.00	&	70.0$\pm$7.5	&	11.7$\pm$0.2	& \\
	& 3.0 (0.02) &		&	182.26$\pm$29.73	&	37.1$\pm$3.4	&	8.4$\pm$0.2	& \\
	& 3.0 (0.02) &		&	89.08$\pm$9.00	&	37.1$\pm$3.4	&	8.4$\pm$0.2	& \\
	& 3.0 (0.02) &	D$_2$CO	&	$<$2.51	&	53	&		& \\
I19035--VLA1 & 4.8 (0.05)	&	H$_2$CO	&	92.65$\pm$2.92	&	60.3$\pm$2.0 &	33.1$\pm$0.1	& 39\\
	& 4.8 (0.05) &	H$_2^{13}$CO	&	4.10$\pm$0.52	&	60.3$\pm$2.0	&	32.9$\pm$0.2	& 39\\
	& 4.8 (0.05) &	HDCO	&	$<$4.79	&	60.3$\pm$2.0	&		& 39\\
	& 4.8 (0.05) &	D$_2$CO	&	$<$3.63	&	60.3$\pm$2.0	&		& 39\\
19410+2336	& 5.8 (0.06)  &	H$_2$CO	&	16.27$\pm$1.34	&	60.1$\pm$4.8	&	22.7$\pm$0.1	& 21\\
	& 5.8 (0.06) &	H$_2^{13}$CO	&	5.59$\pm$0.19	&	25.4$\pm$11.1	&	22.5$\pm$0.1	& 21\\
	& 5.8 (0.06) &	HDCO	&	6.43$\pm$1.82	&	25.4$\pm$11.1	&	22.4$\pm$0.2	& 21\\
	& 5.8 (0.06) &	D$_2$CO	&	1.32$\pm$0.28	&	25.4$\pm$11.1	&	22.6$\pm$0.2	& 21\\
\end{supertabular}
\end{center}

\end{document}